\documentclass{achemso}
\setkeys{acs}{maxauthors=99,etalmode=truncate} 
\usepackage{amsmath}
\usepackage{mathrsfs}
\usepackage{dcolumn}
\usepackage{bm}
\usepackage{multirow}
\usepackage{graphicx}
\usepackage{pdfpages} 
\usepackage[version=3]{mhchem}

\usepackage{mathtools}
\usepackage{braket}

\title{Comparing self-consistent $GW$ and vertex corrected $G_0 W_0$ ($G_0W_0\Gamma$) accuracy for molecular ionization potentials}
\author{Ming Wen}
\author{Vibin Abraham}
\author{Gaurav Harsha}
\affiliation{Department of Chemistry, University of Michigan, Ann Arbor, Michigan 48109}
\author{Avijit Shee}
\affiliation{Department of Chemistry, University of California, Berkeley, California 94720-1460}

\author{K. Birgitta Whaley}
\affiliation{Department of Chemistry, University of California, Berkeley, California 94720-1460}

\author{Dominika Zgid}
\affiliation{Department of Chemistry, University of Michigan, Ann Arbor, Michigan 48109}
\alsoaffiliation{Department of Physics, University of Michigan, Ann Arbor, Michigan 48109}
\email{zgid@umich.edu}

\date{\today}

\begin{document}


\begin{abstract}
We test the performance of self-consistent $GW$ and several representative implementations of vertex corrected $G_0 W_0$ ($G_0W_0\Gamma$). These approaches are tested on benchmark data sets covering full valence spectra (first ionization potentials and some inner valence shell excitations).  For small molecules, when comparing against state of the art wave function techniques, our results show that full self-consistency in the $GW$ scheme either systematically outperforms vertex corrected $G_0 W_0$  or gives results of at least {comparative quality}. 
Moreover, $G_0 W_0\Gamma$ results in additional computational cost when compared to $G_0 W_0$ or self-consistent $GW$. {The dependency of $G_0 W_0\Gamma$} on the starting mean-{field} solution is frequently {more dominant} than the magnitude of vertex correction itself.
Consequently, for molecular systems, self-consistent $GW$ performed on imaginary axis (and then followed by modern analytical continuation techniques) offers a more reliable approach to make predictions of {ionization potentials}. 
\end{abstract}


\section{Introduction}\label{sec:intro}

Accurate and computationally accessible simulation of molecules and solids is a major goal when developing new or improving existing computational approaches.
Wave function methods such as coupled cluster method (CC)~\cite{Cizek_CC,Paldus_Cizek,Bartlett07,gruneisCoupledClusterQuantum2020,schaferLocalEmbeddingCoupled2021,shiManyBodyMethodsSurface2023b} and configuration interaction (CI)~\cite{fosterCanonicalConfigurationalInteraction1960,siegbahnDirectConfigurationInteraction1980,davidsherrillConfigurationInteractionMethod1999} have seen a great success in modeling moderately sized molecules and solids with small unit cells. However, these 
ab-initio simulations for {larger and more complicated systems} still remain extremely challenging, mainly due to the high computational scaling. 

For larger molecules {and} periodic systems, methods with lower computational scaling are favorable. 
Approaches such as density functional theory (DFT) \cite{DFT1,DFT2} are more efficient in tackling systems with large numbers of electrons.
However, these methods often struggle to offer quantitatively accurate predictions when electron correlation plays a significant role. Moreover, for DFT methods there is a lack of a systematic route to improve the  results.

Many-body perturbation theory (MBPT) cast into the Green's function language provides an alternative group of ab-initio methods to model electron correlation.~\cite{fetterQuantumTheoryManyParticle2012}
Most of these methods such as $GW$ \cite{hedinNewMethodCalculating1965, G0W0_Pickett84,G0W0_Hybertsen86,aryasetiawanGWMethod1998,Stan06,Koval14,scGW_Andrey09,GW100,holmFullySelfconsistentMathrmGW1998,QPGW_Schilfgaarde,golzeGWCompendiumPractical2019} or Green's function second order (GF2)~\cite{Snijders:GF2:1990,Dahlen05,Phillips14,Welden16,GF2_Sergei19} can be executed with relatively low computational scaling, especially when compared to wave function theories. 

Since the diagrammatic Green's function expansions are based on perturbation theory, they also hold a promise of being systematically improvable.
In 1965, Hedin proposed a complete set of conjugated equations, detailing the relationships between the Green's function, irreducible polarizability, screened Coulomb interaction, vertex function, and self-energy.~\cite{hedinNewMethodCalculating1965}
The $GW$ method constitutes a simpler truncated formulation of Hedin's equations of MBPT.
{A more approximated} version of $GW$ theory, namely $G_0 W_0$, where only the first iteration of the {self-consistent loop} is performed (vertex function is assumed to be unity), has become one of the most widely applied methodologies in computational electronic structure.
While relatively accurate and computationally inexpensive, $G_0 W_0$ still lacks full quantitative accuracy and can qualitatively fail for {systems with stronger electron correlations}.

One of the proposals of improving $G_0 W_0$ accuracy is the addition of the vertex function $\Gamma$ as a correction term to $G_0 W_0$. Those vertex-corrected methods are commonly referred to as $G_0 W_0\Gamma$.~\cite{Arno1998,romanielloSelfenergyGWLocal2009a,Romaniello2012,maggioGWVertexCorrected2017,wangAssessingApproachHedin2021,ren2015,gruneisIonizationPotentialsSolids2014,kutepovElectronicStructureNa2016}
{Note that in this manuscript, we refer to $G_0 W_0\Gamma$ as a general term for the group of methods that involve the vertex correction. When a specific variant of vertex correction is being discussed, we use a specific name for that given variant such as $G_0W_0\Gamma_X$, $G_0 W_0 \Gamma_0^{(1)}$, or $G_0 W_0 \Gamma^{(\mathrm{NL})}$, introduced later in the Theory section. }

Generally, the vertex correction is expected to address stronger electron correlation (or simply correlation missing at the $G_0 W_0$ level) in both solids and molecules, and improve predictions for the band structure and the photo-electron spectrum.
However, the implementations, approximations, and performances of $G_0 W_0\Gamma$ vary from one version to another because many different strategies can be {used} to lower the computational scaling. 

{The vertex correction will result in addition of new diagrammatic terms to $G_0 W_0$.
Another possible strategy is to avoid adding new diagrams and renormalize the Green's function lines in the $GW$ expressions. 
By performing full self-consistency of the Hedin's equations (excluding the vertex), we can update the Green's function $G$ and self-energy $\Sigma$ in each loop, resulting in a fully self-consistent $GW$ description (sc$GW$).~\cite{kutepovSelfconsistentSolutionHedin2017,Iskakov20,yehFullySelfconsistentFinitetemperature2022a}}

{The formal computational scaling of finite-temperature $GW$ using the Matsubara formalism is $\mathcal O(N^6)$, but it can be reduced to $\mathcal O(N^4)$ when density fitted integrals are employed.~\cite{carusoSelfconsistentGWAllelectron2013a,yehFullySelfconsistentFinitetemperature2022a}} 
When compared with $G_0W_0$, the cost of sc$GW$ differs only by a prefactor {depending on} the number of iterations required to reach convergence.
Multiple variants of the $GW$ self-consistency were introduced in the past. For detailed discussions of these variants, see Refs.~\cite{QPGW_Schilfgaarde,shishkinAccurateQuasiparticleSpectra2007,carusoUnifiedDescriptionGround2012,carusoSelfconsistentGWAllelectron2013a,yehFullySelfconsistentFinitetemperature2022a, Yeh_relativistic, 2C-G0W0-Forster}.

{Consequently, two proposals can be put forward to improve the accuracy of $G_0 W_0$ results: (i) inclusion of the vertex function directly on top of $G_0 W_0$ (the $G_0 W_0\Gamma$ scheme), or alternatively (ii) performing fully self-consistent $GW$ loops (the sc$GW$ scheme). }
We aim to investigate the performance improvement of these two routes upon $G_0W_0$. These comparisons will guide future developments in the Green's function theories towards robust methods that provide a good compromise between low computational scaling and high accuracy. 
{We focus our investigations on small molecules because accurate ionization data from ab-initio methods are readily available.}

{Moreover, we are also interested in this comparison as the reliability of vertex corrections has been a topic of recent discussions in molecular ionization potential (IP) prediction. 
Berkelbach and Lewis noted that the additional diagrams in the $G_0 W_0\Gamma$ scheme did not result in the improvement of the $G_0 W_0$ results, if vertex diagrams were added to the polarizability exclusively, but not to the self-energy.~\cite{lewisVertexCorrectionsPolarizability2019}  
Multiple works also noticed that vertex-corrected $G_0 W_0$ based on Hartree-Fock references did not improve parent $G_0 W_0$ results.~\cite{lewisVertexCorrectionsPolarizability2019, wangAssessingApproachHedin2021, forsterExploringStaticallyScreened2022a} 
However, changing the functional reference to PBE resulted in some improvements reported in in Ref.~\cite{wangAssessingApproachHedin2021}.}
{In the sc$GW$ scheme, the Green's function lines are renormalized in the diagrammatic expansion and the results are not dependent on the initial mean-field.  
}

{In this work,} we employ our recently introduced finite temperature sc$GW$ method~\cite{Iskakov20,yehFullySelfconsistentFinitetemperature2022a, Yeh_relativistic} and compare its results against a {few representative} $G_0W_0\Gamma$ schemes presented earlier in the literature.
We compare against three possible ways of approximating vertex, (i) stochastic methodology developed in the group of Vl\v{c}ek in Ref.~\cite{mejuto-zaeraAreMultiquasiparticleInteractions2021}, (ii) $G_0W_0\Gamma_0^{(1)}$ implemented by Wang et al. and presented in Ref.~\cite{wangAssessingApproachHedin2021}, and (iii) $G_0W_0\Gamma^{\mathrm{(NL)}}$ implemented in the Kresse's group and presented in Ref.~\cite{maggioGWVertexCorrected2017} by Maggio \textit{et al.}

sc$GW$ and vertex corrected $G_0 W_0$ methodologies are tested on molecular examples {and} validated against wave function theories or {experimental data}.
We examine these approaches both for the first and inner valence shell IPs {depending on the availability of related data sets}. {Note that the respective performance of sc$GW$ and vertex-corrected $GW$ on electronic gases~\cite{holmFullySelfconsistentMathrmGW1998} and metals~\cite{takadaInclusionVertexCorrections2001a} can differ from molecular cases. In this work, we restrict our discussion to the molecular regime unless specified otherwise. } 

We present our work in this article as follows. 
In the Theory section, we provide a concise introduction to finite temperature sc$GW$ and a brief review of recent developments in adding the vertex correction to $G_0W_0$.
The Computational Procedure section describes the general calculation protocol we used to conduct predictions.
The Results section contains findings of this work.
We summarize our findings and outlooks in the Conclusions section.

\section{Theory}\label{sec:theory}

\subsection{Hedin's equations}
A set of self-consistent equations detailing the relationship among the self-energy, Green's function, screened Coulomb interaction, and irreducible polarizability in many-electron systems was formulated in Hedin's foundational article.~\cite{hedinNewMethodCalculating1965}

In the derivation of Hedin's equations, a bare Coulomb interaction $v$ is affected (or ``screened") by the many-electron environment. The screened Coulomb interaction $W$ is defined as
\begin{equation}
    W(12) = v(12) + W(13)P(34)v(42), 
\end{equation}
where  numeric compact indices are employed as a shorthand to represent the state of space-time and spin as $1 = (\textbf{x}_1,\sigma_1,t_1$). Integration over repeated indices is assumed. 
The irreducible polarizability is defined as
\begin{equation}
    P(34) = iG(45)G(64)\frac{\delta G^{-1}(56)}{\delta V(3)},
\end{equation}
where $V$ is the time evolution operator. 
The functional derivative ${\delta G^{-1}(56)}/{\delta V(3)}$ is called the vertex function, and expressed as
\begin{equation}
\begin{split}
\label{eq: vertex}
    \Gamma(12;3) {}&\equiv -\frac{\delta G^{-1}(12)}{\delta V(3)} = \delta(12)\delta(13) + \frac{\delta \Sigma(12)}{\delta V(3)} \\
    &=\delta(13) \delta(23)+\frac{\delta \Sigma (12)}{\delta G(45)} G(46) G(75) \Gamma(67;3).
\end{split}
\end{equation}
The inclusion of vertex function can be viewed as the information about the interaction {among} electrons and holes.~\cite{minnhagenVertexCorrectionCalculations1974,onidaElectronicExcitationsDensityfunctional2002}

\begin{figure}
    \centering
    \includegraphics[scale=0.46]{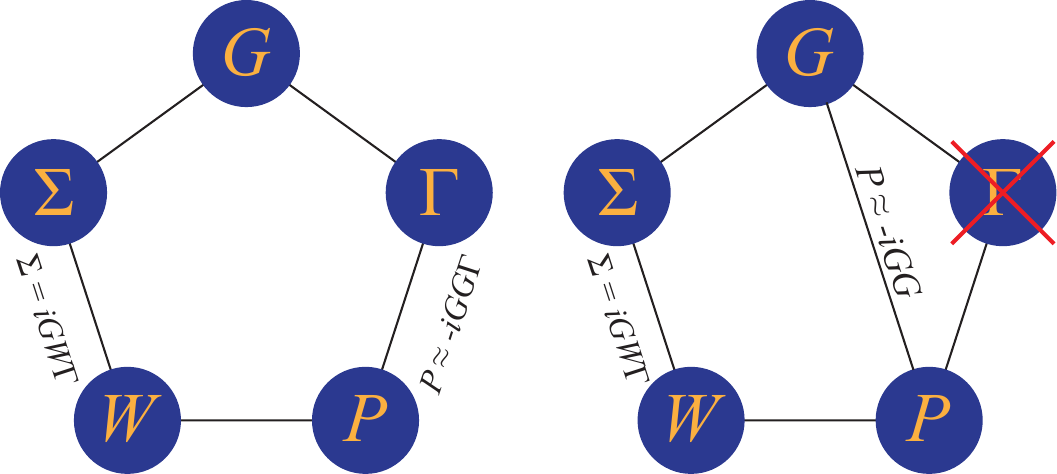}
    \caption{Left: The self-energy, Green's function, vertex function, irreducible polarizability, and screened Coulomb interaction conjugated by Hedin's equations. Right: The $GW$ approximation loop of Hedin's equations without inclusion of the vertex function.}
    \label{fig:Hedin's Pentagon}
\end{figure}

In the original formulation, evaluating the exact vertex function significantly increases the computational cost because it is difficult to calculate a four-index functional derivative.
If the vertex function {is} truncated at the zeroth order in Eq.~(\ref{eq: vertex}) as $\Gamma(12;3) \approx \delta(12)\delta(13)$, then the self-energy $\Sigma$ and the reduced polarizability $P$ can be calculated without evaluating the functional derivative as
\begin{equation}
\begin{split}
\label{eq: self-energy}
\Sigma(12) {}&= iW(13)G(14)\Gamma(42;3)\\
& \approx iG(12)W(12),
\end{split}
\end{equation}

\begin{equation}
\begin{split}
\label{eq: polarizability}
P(12) {}&= -iG(23)G(42)\Gamma(34;1)\\
& \approx -iG(12)G(21).
\end{split}
\end{equation}
The name $GW$ approximation (GWA) was coined to signify the exclusion of vertex function in Hedin's equations.\cite{hedinNewMethodCalculating1965,aryasetiawanGWMethod1998}
The GWA scheme is depicted by the trapezoidal loop in the right diagram of Figure~\ref{fig:Hedin's Pentagon}.

\subsection{$G_0 W_0$ and self-consistent $GW$}
GWA can be executed in two ways. One of them called  $G_0 W_0$ is the first iteration of the Hedin's equations loop without the inclusion of the vertex function where the zeroth order Green's function $G_0$ is evaluated using a mean-field Hamiltonian (Hartree-Fock or DFT).
{Then the Green's function is used to evaluate the self-energy.}
Usually, only the diagonal elements of the self-energy matrix are evaluated and the quasiparticle (QP) approximation is then employed to afford spectral quantities.~\cite{golzeGWCompendiumPractical2019}

Fully self-consistent $GW$ (sc$GW$) is executed if {all the quantities (excluding the vertex function) described by Hedin's equations} are iterated until self-consistency.
For the implementation details of sc$GW$ as performed in the Zgid group, we encourage the reader to consult Ref.~\cite{Tran_GW_SEET} for molecular problems, Refs.~\cite{Iskakov20,yehFullySelfconsistentFinitetemperature2022a,Yeh_MnO3} for periodic solid problems, and Ref.~\cite{Yeh_relativistic} for relativistic problems in periodic solids.

In our finite temperature sc$GW$ implementation, we express the frequency/time dependent quantities on an imaginary axis grid.~\cite{Kananenka16,Iskakov_Chebychev_2018,dong2020legendrespectral,liSparseSamplingApproach2020}
No diagonal approximation to the self-energy is used and all the matrix elements of self-energy for all the frequencies are evaluated and included in the self-consistent equations. 
The total self-energy in sc$GW$ is expressed as
\begin{equation}
\Sigma^{GW}[\mathcal{G}](i\omega)=
{\Sigma}_{\infty}^{GW}[\mathcal{G}] + \tilde {\Sigma}^{GW}[\mathcal{G}](i\omega),
\end{equation}
where ${\Sigma}_{\infty}^{GW}[\mathcal{G}]$ is the static part of self-energy evaluated using the first order diagrams (Hartree and exchange) employing the correlated one-body density matrix, and $\tilde {\Sigma}^{GW}[\mathcal{G}](i\omega)$ is the frequency-dependent dynamical part of self-energy.  
Both self-energy parts, dynamic and static, depend on Green's function and are evaluated iteratively until achieving self-consistency. 

{In our implementation, the density-fitting approximation~\cite{dunlapRobustVariationalFitting2000} is used to decompose two-electron integrals. Overall, the computational cost of our sc$GW$ scales as  $\mathcal O (N_\tau N_{\mathrm{orb}}^2 N_{\mathrm{aux}}^2)$. where $N_\tau$ is the number of imaginary time grid points, $N_{\mathrm{orb}}$ is the number of orbitals in the problem, and $N_{\mathrm{aux}}$ is the number of auxiliary basis functions used in the density fitting procedure. }

No QP approximation is evoked to evaluate spectral quantities in our version of sc$GW$. To yield spectral information, the finite-temperature Green's function from the imaginary axis is continued to the real frequency axis with the help of the Nevanlinna analytical continuation technique introduced by 
Fei \textit{et al.}  \cite{feiNevanlinnaAnalyticalContinuation2021,huangRobustAnalyticContinuation2023}
The {spectral function} can be derived from the continued Green's function as
\begin{equation}
    \mathcal{G}(i\omega) \xrightarrow[\mathrm{analytical\ continuation}]{\mathrm{Nevanlinna}} A(\omega) = -\frac{1}{\pi}\mathrm{Im}\left[\mathrm{Tr[}G(\omega)]\right].
\end{equation}

Spectral functions rendered by Nevanlinna analytical continuation are positive and normalized by definition, which helps to resolve the isolated ionization peaks in the PES. 
{We use the notation ``sc$GW$'' to refer to our finite-temperature implementation which exactly follows the procedure of self-consistent Hedin equation without including the vertex. This procedure is then followed by the Nevanlinna analytical continuation. 
Our sc$GW$ scheme should contrasted to other self-consistency schemes such as the ones proposed by Schilfgaard-Kotani-Faleev ~\cite{faleevAllElectronSelfConsistentGW2004,shishkinSelfconsistentGWCalculations2007}, and eigenvalue-self-consistent $GW$.~\cite{vanschilfgaardeQuasiparticleSelfConsistentGW2006} These schemes constitute a further departure from Hedin's original scheme.}

\subsection{Practical implementations of vertex correction}

It is believed that introducing the vertex function $\Gamma$ as a correction term on top of {GWA} can improve its accuracy.
The rigorous evaluation of vertex function is difficult due to the presence of the four-point functional derivative ($\delta \Sigma/\delta G$) in Eq.~\ref{eq: vertex} and due to the necessity of iterative self-consistent evaluation of 
 both self-energy $\Sigma$ and irreducible polarizability $P$ since they involve $\Gamma$ in their respective Eqs.~\ref{eq: self-energy} and \ref{eq: polarizability}. 
Consequently, in practical implementations of the vertex correction, many approximations are introduced to lower the computational cost and make such evaluations viable.

It is frequently argued that sc$GW$  overestimates band gaps.~\cite{holmFullySelfconsistentMathrmGW1998, caoFullyConvergedPlanewavebased2017} Thus $G_0W_0$, which benefits from the error cancellation between the {mean-field} starting point and a subsequent $GW$ evaluation, is used as an affordable alternative. 
In this approximation, 
$\Gamma$ is evaluated on top of $G_0W_0$ resulting in a $G_0W_0\Gamma$ method.
However, even in this simplified scenario further approximations are necessary to make the final evaluation of $G_0W_0\Gamma$  possible.
Here, we summarize recent developments of practical $G_0W_0\Gamma$ implementations.

Reining and coworkers introduced the $\rho/G$-approach, \cite{delsoleGWEnsuremathGamma1994,brunevalManyBodyPerturbationTheory2005, romanielloSelfenergyGWLocal2009a} in which a two-part vertex correction was proposed as
\begin{equation} \label{eq:reiningv}
\begin{split}
\Gamma(12;3) ={}& \delta(13) \delta(23) + \delta(12) f_{\mathrm{xc}}^{\mathrm{eff}}(14)P(43)\\ 
&+ \Delta \Gamma(12;3),
\end{split}
\end{equation}

\begin{equation}
\begin{split}
\Delta \Gamma(12;3) = \left[\frac{\delta\Sigma_{\mathrm{xc}}(12)}{\delta \rho (4)} - \delta(12)f_{\mathrm{xc}}^{\mathrm{eff}}(14) \right]P(43).
\end{split}
\end{equation}
The first two terms in Eq.~\ref{eq:reiningv} are defined as a ``local" part while $\Delta \Gamma$ is called the ``non-local" term because $\Delta \Gamma$ has zero contribution to $P$. 

In both the local and non-local parts, $f_{\mathrm{xc}}^{\mathrm{eff}}$ is included. 
It serves as an auxiliary effective function used to obtain $P$ easily from only two-point quantities.
$f_{\mathrm{xc}}^{\mathrm{eff}}$ has an exact definition which involves a three-point kernel $\delta\Sigma_{\mathrm{xc}}/\delta \rho$, but is often approximated with $v_{\mathrm{xc}}$ ($\delta\Sigma_{\mathrm{xc}}/\delta \rho \approx \delta v_{\mathrm{xc}}/\delta\rho$) retrieved from the starting mean-{field} calculation. 
Such a local vertex correction is relatively inexpensive to compute, but alone it is insufficient in improving $G_0 W_0$ accuracy.~\cite{morrisVertexCorrectionsLocalized2007a,hungExcitationSpectraAromatic2016,delsoleGWEnsuremathGamma1994}

Kresse and co-workers~\cite{starkeSelfconsistentGreenFunction2012a,maggioGWVertexCorrected2017} reported their version of {non-local vertex correction} implementation with the re-formulated four-point notation, given by
\begin{equation} \label{}
\begin{split}
\Gamma(1234) ={}&\delta(13)\delta(24)\\
&+i\frac{\delta\Sigma_{\mathrm{xc}}(12)}{\delta G(56)} G(57) G(86) \Gamma(7834),
\end{split}
\end{equation}
instead of the Hedin's original three-point one. 
The four-point kernel is approximated as
\begin{equation} \label{}
\begin{split}
i\frac{\delta\Sigma_{\mathrm{xc}}(12)}{\delta G(56)} \approx -\delta(26) \delta(15) \delta(t_1,t_2)W^{\mathrm{RPA}}(\textbf{x}_1,\textbf{x}_2,\omega=0),
\end{split}
\end{equation}
where $W^{\mathrm{RPA}}$ is the screened Coulomb interaction, estimated with the random phase approximation employing the frequency-dependent exact-exchange kernel $f_{\mathrm x}$ (RPAx).~\cite{szaboInteractionEnergiesClosedshell1977,colonnaMolecularBondingRPAx2016}
Moreover, the static estimation of $\omega=0$ is applied,~\cite{watabeInfluenceCoulombCorrelation1963,hedinNewMethodCalculating1965} assuming that the kernel is relatively frequency independent. 
Even with RPAx, it is still preferred to evaluate the approximated kernel with a HF starting point instead of a $GW$ reference state. By doing so, screened Coulomb interaction $W^{\mathrm{RPA}}$ could be further simplified as the bare Coulomb interaction $v(\textbf{x}_1,\textbf{x}_2)$ as
\begin{equation} \label{}
\begin{split}
i\frac{\delta\Sigma_{\mathrm{xc}}(12)}{\delta G(56)} \approx -\delta(26) \delta(15) \delta(t_1,t_2)v(\textbf{x}_1,\textbf{x}_2).
\end{split}
\end{equation}
The implementation of $G_0W_0\Gamma$ scales formally as $\mathcal{O}(N^6)$.~\cite{maggioGWVertexCorrected2017}
{In this work we refer to this implementation as $G_0W_0\Gamma^{(\mathrm{NL})}$, where ``NL'' stands for ``non-local''.}

Wang \textit{et al.}~\cite{wangAssessingApproachHedin2021} reported vertex function truncated at the first order. By plugging in $GW$ approximated self-energy $\Sigma^{GW}$ in the $\delta \Sigma/\delta G$ kernel as
\begin{equation} \label{}
\begin{split}
\frac{\delta \Sigma(12)}{\delta G(45)} {}& \approx \frac{\delta \Sigma^{GW}(12)}{\delta G(45)} = \frac{\delta [iG(12)W(12)]}{\delta G(45)} \\
& = i\delta(14)\delta(25)W(12)+G(12)\frac{\delta W(12)}{\delta G(45)}.
\end{split}
\end{equation}
Inserting the approximated expression into Eq.~(\ref{eq: vertex}), it becomes
\begin{equation} \label{}
\begin{split}
\Gamma(12;3) = {}&\delta(13) \delta(23)\\
& + iW(12)G(16)G(72)\underbrace{\Gamma(67;3)}_{\approx\delta{(63)}\delta{(73)}} + \dots 
\end{split}
\end{equation}
By truncating any terms involving order equal to or higher than $\mathcal{O}(W^2)$, the first order vertex function is equivalent to the full second-order self-energy in terms of $W$ (FSOS-$W$) as
\begin{equation} \label{}
\begin{split}
\Gamma^{(1)}(12;3)\approx{}&\delta(13) \delta(23)\\
& + iW(12)G(13)G(32).
\end{split}
\end{equation}

Since only a single iteration is {performed to evaluate such a vertex}, a subscript of 0 is added to $\Gamma^{(1)}$.
The computation cost for $G_0W_0\Gamma_0^{(1)}$ formally scales as $\mathcal{O}(N^{5})$.~\cite{wangAssessingApproachHedin2021}

Vlček~\cite{vlcekStochasticVertexCorrections2019} utilized the non-local exchange term in the starting mean-field calculation for the $G_0 W_0$ method to construct an approximated non-local vertex function $\Gamma_X$. 
In this work, the four-point derivative kernel is approximated as
\begin{equation}
\begin{split}
\frac{\delta \Sigma_{\mathrm{total}}(12)}{\delta G(45)} {}& \approx \frac{\delta [\Sigma_{\mathrm{hartree}}(12)+\Sigma_{\mathrm{exchange}}(12)]}{\delta G(45)} \\
& = -\nu(25)\delta(45)\delta(12) + \nu(12)\delta(52)\delta(41),
\end{split}
\end{equation}
where $\nu(12) = \frac{\delta(t_1-t_2)}{|r_1-r_2|}$.
In the evaluation process of $\Gamma_X$, a stochastic sampling method, instead of the commonly used deterministic one, is used to further minimize computational cost, which gives the evaluation of $\Gamma_X$ a sub-linear computational scaling.~\cite{vlcekStochasticVertexCorrections2019}
{We refer to this approach simply as $G_0W_0\Gamma_X$ for clarity, but readers should note its stochastic nature.}

To summarize, the approximations often used to implement vertex corrections can be categorized by five major categories:
(i) estimation of $\delta \Sigma/\delta G$ kernel; (ii) truncation of $\Gamma$, or approximation of $\Gamma$ with diagrammatic approaches; (iii) correction of $\Sigma$ only (no correction of $P$); (iv) operation only on the diagonal elements of certain matrices; (v) avoidance of full self-consistency. 
Even though all these variants are under the same name of ``vertex corrections", their practical implementations can be significantly different from one to another. 

{In this article, we compare the performance of our sc$GW$ with $G_0W_0\Gamma_X$, $G_0W_0\Gamma_0^{(1)}$, and $G_0W_0\Gamma^{(\mathrm{NL})}$.}

\section{Computational Procedure}\label{sec:comp_proc}

Geometries for molecules within the $GW$100 set~\cite{vansettenGW100BenchmarkingG0W02015a} were obtained from Ref.~\cite{knightAccurateIonizationPotentials2016}. 
Any additional molecules not included the $GW$100 set were selected from the G2 data set.~\cite{curtiss_assessment_1997} The {additional} G2 geometries were taken from Ref.~\cite{haunschildNewAccurateReference2012}.

All sc$GW$ calculations were performed on the imaginary time and frequency axes using our implementation reported in Ref.~\cite{yehFullySelfconsistentFinitetemperature2022a}. For the starting mean-field calculations, we use both HF and DFT (PBE functional).~\cite{perdewGeneralizedGradientApproximation1996} We employed the inverse temperature of $\beta = 1000\ E_\mathrm{h}^{-1}$, which corresponds to temperature $T = 315.8\ \mathrm{K}$. 
To render the spectral function from our results on the imaginary frequency axis, converged Green's functions from sc$GW$ are continued to the real frequency axis with Nevanlinna analytical continuation technique.~\cite{feiNevanlinnaAnalyticalContinuation2021,huangRobustAnalyticContinuation2023} 
{This procedure allows us to identify the first and inner valence shell IP values from the spectral function peaks.}

In some cases, we provide the complete basis set (CBS) limit~\cite{cbs1,CBS2} using our sc$GW$ {results}. In such cases, we extrapolate the results with respect to the inverse of the numbers of orbitals that vary with the basis set size. The function is { $f(n) = E^{\mathrm{IP}}_{\infty} + kn^{-1}$, in which $E^{\mathrm{IP}}_{\infty}$ is the extrapolated IP value and $n$ is the number of basis functions.~\cite{rangelEvaluatingGWApproximation2016}}

{For the $G_0W_0$ calculations that are reported from our code, we follow the same imaginary axis methodology. We transform the first iteration of sc$GW$ self-energy (on the Matsubara axis) to the molecular orbital basis corresponding to the initial mean-field solution (HF or PBE).}
We then use the Pad\'e analytic continuation,~\cite{hanAnalyticContinuationPade2017} implemented in the \texttt{PySCF} quantum chemistry package,~\cite{sunPySCFPythonbasedSimulations2018,sunRecentDevelopmentsPySCF2020} to obtain the self-energy on the real frequency axis.
Ionization potential and other quasiparticle excitations are then calculated by solving the quasiparticle equation,
\begin{equation}
    \epsilon_p = \epsilon_p^0 + \braket{\phi_p | \left(\Sigma (\epsilon) - v_{xc} \right)| \phi_p},
\end{equation}
where $\braket{\phi_p | \Sigma(\omega) | \phi_p}$ is the analytically continued diagonal part of the self-energy, and $v_{xc}$ is the Hartree plus exchange-correlation potential of the corresponding mean-field reference.
{$G_0W_0\Gamma$ results and statistical criteria in the Results section are either cited or or calculated from data reported in relevant literature.}

Other input parameters for our calculations (basis set, starting mean-field, \textit{etc.}) vary to facilitate  different comparisons. See corresponding subsections in the Results section and Supporting Information for a more detailed description.

\section{Results}\label{sec:results}

We employ the finite-temperature sc$GW$ method to determine molecular IPs and conduct a comparative analysis of its performance against previously established vertex-corrected $GW$ results.
{To ensure that $G_0 W_0$ and sc$GW$ from our code can be directly compared with the $GW$100 benchmark data, we first inspect the validity of our finite-temperature $GW$ by comparing it with past $GW$ implementations in Fig. S4 of the Supporting Information. This test demonstrates that the density fitting procedure used in our $GW$ implementation, and necessary analytical continuation techniques to obtain spectral data, have no significant bearing on accuracy of our data when compared against the $GW$100 benchmark data.}

{In this section, we first compare our sc$GW$ results and the valence shell excitations obtained for a stochastic implementation of $G_0W_0\Gamma_X$.~\cite{mejuto-zaeraAreMultiquasiparticleInteractions2021}
We then move on to focus on only the first IP peaks for which more data are available.
To this end, we compare the first-order vertex corrected $G_0W_0\Gamma_0^{(1)}$ method with our sc$GW$ results for the $GW$100 dataset.~\cite{wangAssessingApproachHedin2021}
Similarly, we compare sc$GW$ to $G_0W_0\Gamma^{(\mathrm{NL})}$ for another set of 29 molecules.~\cite{maggioGWVertexCorrected2017}}
These results are compared with either highly accurate theoretical reference or experimental benchmark data.

\subsection{Comparison of sc$GW$ and stochastic $G_0W_0\Gamma_X$ for valence shell excitations}\label{subsec:valence}

The objective of this subsection is to compare the IP peaks obtained from our sc$GW$ implementation to stochastic $G_0W_0\Gamma_X$ as implemented and presented by Vlček and co-workers.~\cite{mejuto-zaeraAreMultiquasiparticleInteractions2021}
To do so, for five small systems, we look at the IP peaks arising from both the first (or the highest occupied molecular orbital, i.e. HOMO) and the {inner valence shell} excitations: H$_2$O, N$_2$, NH$_3$, C$_2$H$_2$, CH$_4$ and three additional systems:  CO, HF and C$_2$H$_4$.

To establish an equal footing for comparing the two methods, we use the same basis sets, extrapolation technique and molecular geometries as in Ref.~\cite{mejuto-zaeraAreMultiquasiparticleInteractions2021}.
That is, we perform calculations in aug-cc-pVXZ (X = D, T, Q) basis sets~\cite{aug-cc-H-B-Ne,aug-cc-Al-Ar,EMSL-paper-new} and use HF data as initial inputs for the $GW$ calculations. 
Final sc$GW$ results are extrapolated to the basis set limit in a given basis set family.~\cite{rangelEvaluatingGWApproximation2016}
We report the first few IP peaks for each molecule so that non-degenerate isolated energy levels can be clearly differentiated. 
Results from both sc$GW$ and stochastic $G_0W_0\Gamma_X$ are compared against reference data from adaptive sampling configuration interaction (ASCI)~\cite{tubmanDeterministicAlternativeFull2016, holmesHeatBathConfigurationInteraction2016} method, also reported in Ref.~\cite{mejuto-zaeraAreMultiquasiparticleInteractions2021}.

\begin{figure*}[]
    \centering
    \includegraphics[scale=0.65]{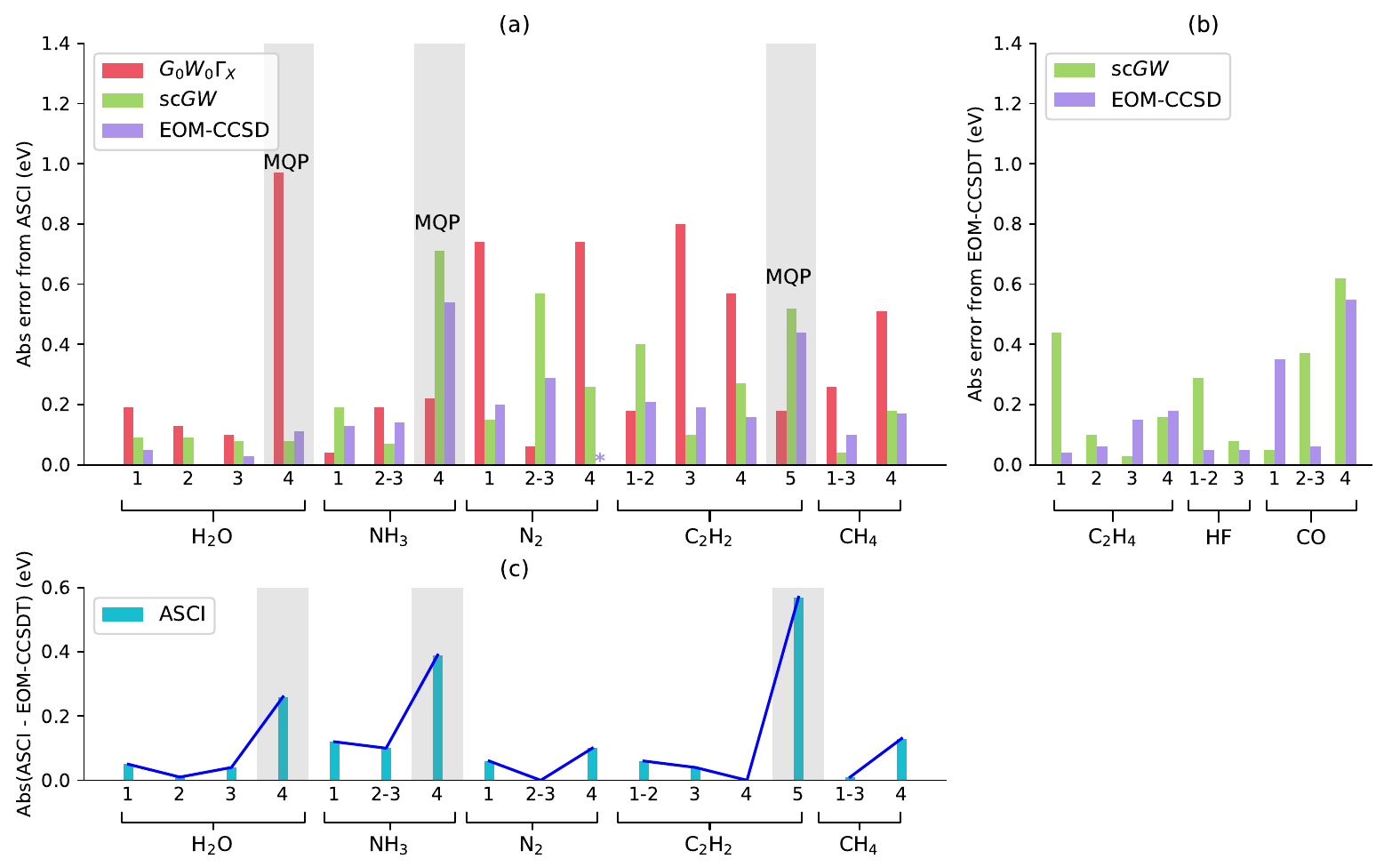}
    \caption{(a) Absolute errors of sc$GW$, $G_0W_0\Gamma_X$,~\cite{mejuto-zaeraAreMultiquasiparticleInteractions2021} and EOM-CCSD predictions of the first and inner shell ionization peaks as compared with ASCI benchmarks. Note that the bar signified with a purple ``*" sign means that EOM-CCSD calculation did not converge and the absolute error is not necessarily zero. Shaded areas are within the multi-quasiparticle peak (MQP) regime. (b) Additional molecules calculated with sc$GW$ and EOM-CCSD compared with EOM-CCSDT benchmarks. (c) Absolute errors between ASCI and EOM-CCSDT benchmarks.}
    \label{fig:MAE inner}
\end{figure*}

In Fig.~\ref{fig:MAE inner}(a), we plot IP peaks computed by stochastic $G_0W_0\Gamma_X$, sc$GW$ and EOM-CCSD compared against ASCI benchmarks for the 16 excitations from the 5 systems.
Out of the total 16 values reported, sc$GW$ yields more accurate data than $G_0W_0\Gamma_X$ for 11 peaks. 
However, out of the five peaks where $G_0W_0\Gamma_X$ is better than sc$GW$, only for the following three peaks, {the difference is} more than 0.2 eV: 
the fourth ionization peak of ammonia, the second/third peaks of N$_2$, and the fifth peak of acetylene. 
For the IP data in Fig.~\ref{fig:MAE inner}(a), the mean {absolute} error (MAE) with respect to ASCI for sc$GW$ is {0.24} eV, and for stochastic $G_0W_0\Gamma_X$ is 0.37 eV.
Compared with a 1.47 eV {MAE} given by $G_0W_0$, both sc$GW$ and $G_0W_0\Gamma_X$ yield inner excitations with much better accuracy.

In addition to the ACSI method, we also compare our results to equation of motion coupled cluster (EOM-CC)~\cite{geertsenEquationofmotionCoupledclusterMethod1989,stantonEquationMotionCoupledcluster1993,krylovEquationofMotionCoupledClusterMethods2008,musialEquationofmotionCoupledCluster2003,ranasingheVerticalValenceIonization2019} hierarchy.
EOM-CC methods are systematically improvable when adding more excitations and hence here we use EOM-CCSD and EOM-CCSDT values.
All EOM-CC calculations were performed using \texttt{CFOUR} quantum chemistry package.~\cite{cfourJCP2020} 
Both EOM-CCSD and EOM-CCSDT were performed with aug-cc-pVQZ basis and not extrapolated to the basis set limit. 
In Fig.~\ref{fig:MAE inner}(c), we look at the difference between inner IPs predicted by EOM-CCSDT and ASCI benchmarks.
We observe that the two methods are generally in good agreement with each other (see Table. SI
in the Supporting Information).
Therefore, EOM-CCSDT could also be used as a benchmark against which $GW$ results can be compared.
Discrepancies between ASCI and EOM-CCSDT tend to increase for deep inner excitations, it remains unclear which of the two methods is more accurate. 

Moreover, the comparison allows us to assess the sc$GW$ method against different levels of excitations used in EOM-CC, i.e. EOM-CCSD and EOM-CCSDT.
This is helpful for understanding the accuracy that can be expected from sc$GW$ for IP predictions.

In Fig.~\ref{fig:MAE inner}(b), we compare the ionization peaks from sc$GW$ and EOM-CCSD for ethylene (C$_2$H$_4$), hydrogen fluoride (HF) and carbon monoxide (CO).
For these systems, we employ EOM-CCSDT results in the aug-cc-pVQZ basis-set as reference.
Overall, based on the results in panels (a) and (b) of Fig.~\ref{fig:MAE inner}, as well as additional data in Fig. S3 of Supporting Information, we deduce that sc$GW$ is similar in accuracy as EOM-CCSD.
This similarity is not surprising as connections between $GW$, RPA, and coupled cluster theory have been well studied.~\cite{scuseria_particle-particle_2013,mcclain_spectral_2016,lange_relation_2018,quintero-monsebaiz_connections_2022,tolle_exact_2023}
However, it is worth noticing that EOM-CCSD data were difficult to obtain due to convergence problems for some of the inner peaks while sc$GW$ converged without any difficulty.
It is also worth mentioning that for the 4th peaks of H$_2$O and N$_2$, $G_0W_0\Gamma_X$ gave relatively poor results. For H$_2$O and N$_2$, sc$GW$ gave very good results, confirming that the difficulty in illustrating these IPs comes from lack of optimization of orbitals in $G_0W_0\Gamma_X$ and not necessarily from the presence of strong correlation. 
Only the 4th peak of nitrogen displays signs of strong correlation which is not recovered by $G_0W_0\Gamma_X$, and also in EOM-CCSD due to lack of complete convergence.

Further analyzing the results in Fig.~\ref{fig:MAE inner}, the IP data, particularly in panel (a), can be separated into two regimes: (i) single, individual quasiparticle (SQP) and (ii) multi-quasiparticle (MQP).~\cite{cederbaumManyBodyEffectsValence1980,mejuto-zaeraAreMultiquasiparticleInteractions2021}
For SQPs, the quasiparticle (or the ionization) peak carries most of the spectral weight. Such peaks can be recovered accurately by $GW$ methods.~\cite{guzzoValenceElectronPhotoemission2011}
{On the other hand, for MQP, a significant amount of spectral weight is transferred to satellite features, often referred to as shake-up satellites in molecules.~\cite{mejuto-zaeraAreMultiquasiparticleInteractions2021}
Mean-field and perturbative methods such as DFT and $GW$ are not adept in describing MQPs, as one needs to account for complicated interactions among many excited states.\cite{onidaElectronicExcitationsDensityfunctional2002}
The general belief is that one can obtain better results for the MQPs by adding higher-order quantum corrections via inclusion of the vertex term, which describes dynamical two-particle correlations.\cite{shishkinAccurateQuasiparticleSpectra2007}}
However, when looking at the accuracy of the fully self-consistent $GW$ in Fig.~\ref{fig:MAE inner}, we observe that both peaks in the SQP and MQP regimes are well recovered by sc$GW$.
Therefore, at least based on the examples considered here, we can confidently say that achieving full self-consistency in $GW$ provides results that are equally, if not more, accurate than $G_0W_0\Gamma$.
While both methods provide significant improvements over $G_0W_0$, one could argue that sc$GW$ is more advantageous over $G_0W_0\Gamma$ as it provides access to other thermodynamic quantities such as total energy, entropy, \textit{etc.}, in addition to the reliable IP prediction and it is independent of the starting point.

Consequently, at least for the examples listed in Fig.~\ref{fig:MAE inner}(a), the vertex correction seem to be unnecessary and very good results can be obtained by employing sc$GW$ alone.
To confirm the points observed above, we plotted spectral functions for selected molecules in the Supporting Information (see Fig. S1 and S2).
We observe that sc$GW$ also produces similar complicated MQP regime peak structures as reported in Ref.~\cite{mejuto-zaeraAreMultiquasiparticleInteractions2021}.

\subsection{Comparison of sc$GW$ and $G_0W_0\Gamma_0^{(1)}$ for first IP peaks}\label{subsec:gw100rinke}

Here, we compare the performance of our sc$GW$ to $G_0W_0\Gamma^{(1)}$ as implemented by Wang et al. and presented in Ref.~\cite{wangAssessingApproachHedin2021} for the $GW$100 dataset.\cite{vansettenGW100BenchmarkingG0W02015a,Krause2015,GW100,Forster2021} 
This comparison is done only for the first ionization potential peaks, where molecular data sets are more readily available. 
To ensure that our sc$GW$ data can be compared against the $G_0W_0\Gamma^{(1)}$ data , we evaluated our sc$GW$ in the same basis set, \textit{i.e.}, def2-TZVPP. For similar reason, for $G_0W_0$ data, we compare both Hartree-Fock and PBE as mean-field starting points for $GW$ calculations. 

\begin{figure*}
\includegraphics[scale=0.68]{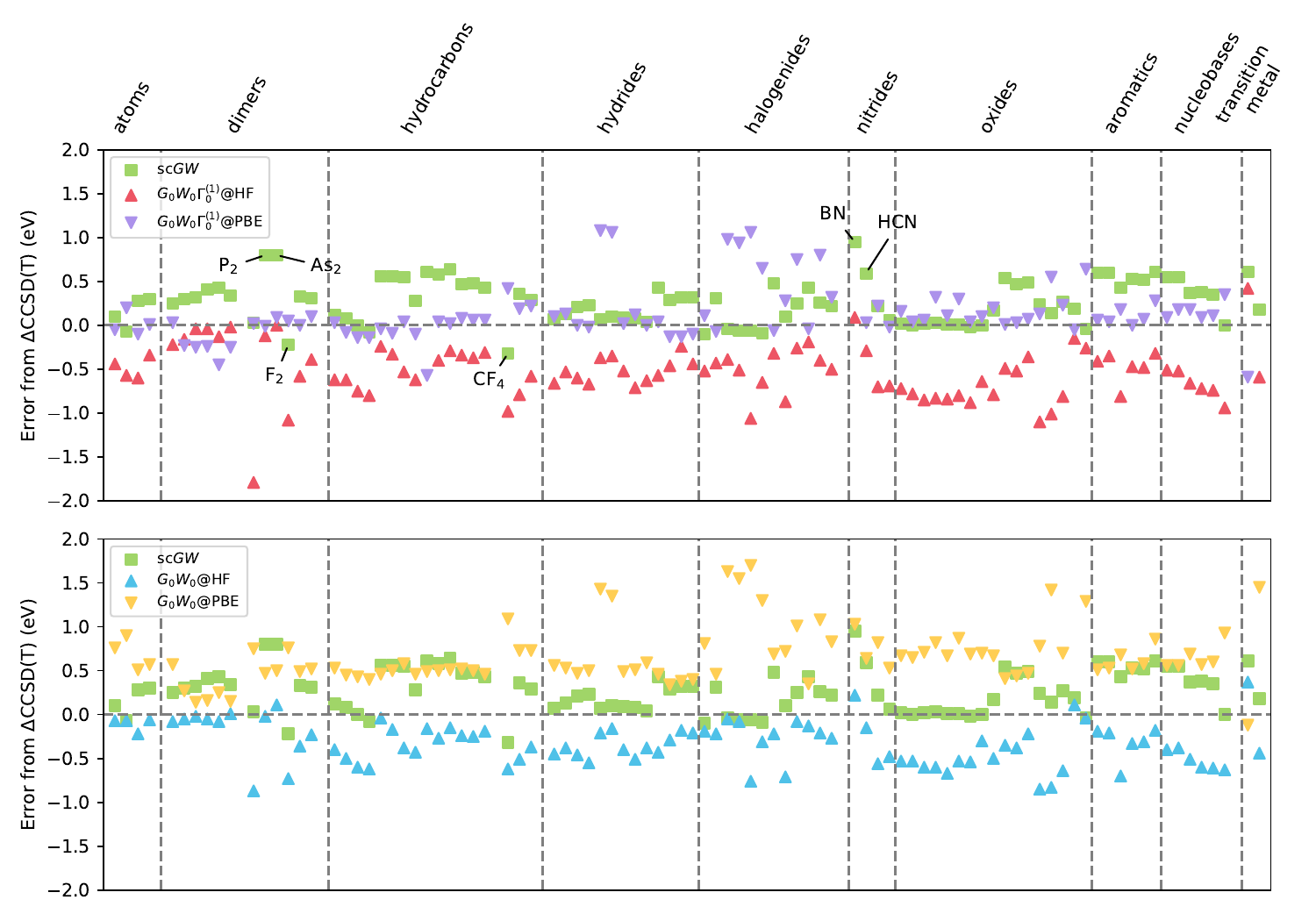}
\centering
\caption{Signed errors from $\Delta$CCSD(T) benchmarks, of $G_0W_0$,  sc$GW$, and $G_0W_0\Gamma_0^{(1)}$~\cite{wangAssessingApproachHedin2021} for the $GW$100 data set. Some outliers (within their own group) are labeled. Top panel: performances of sc$GW$ and $G_0W_0\Gamma_0^{(1)}$ with different starting mean-field calculations. Bottom panel:  performance of sc$GW$ and $G_0 W_0$ with different starting mean-field calculations.
}
\label{fig:Dev GW100} 
\end{figure*}

\begin{figure*}
\includegraphics[scale=0.55]{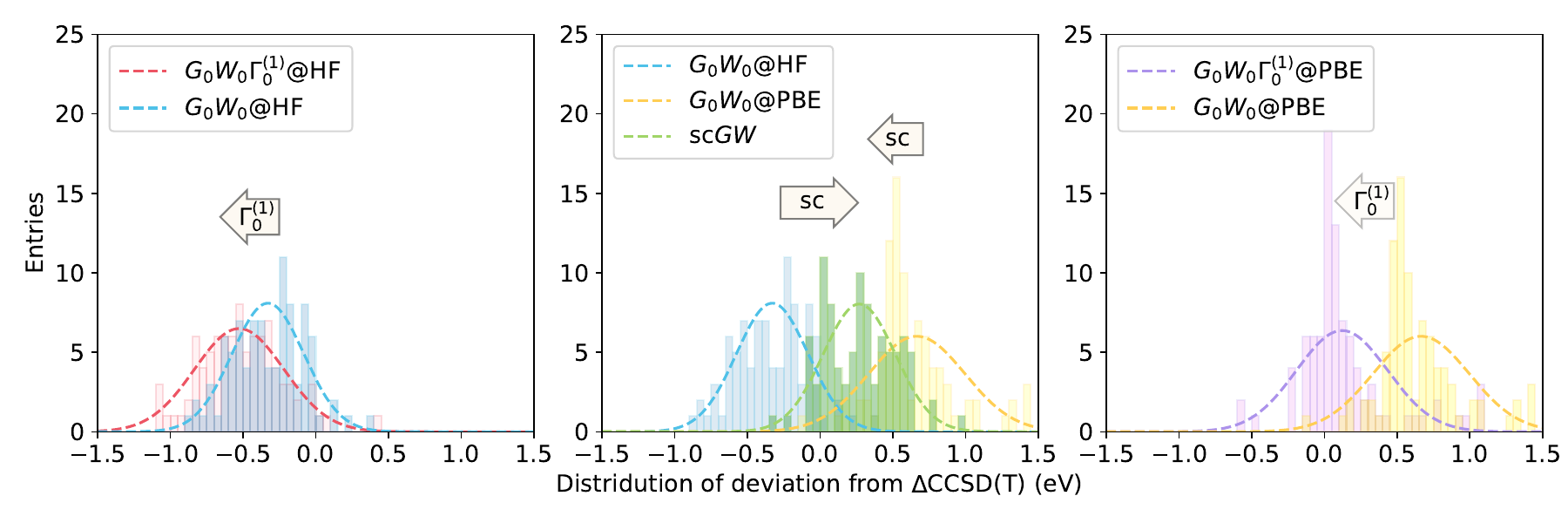}
\centering
\caption{Signed errors from $\Delta$CCSD(T) benchmarks, of $G_0W_0$,  sc$GW$, and $G_0W_0\Gamma_0^{(1)}$~\cite{wangAssessingApproachHedin2021} for the $GW$100 data set. Dashed curves are fitted Gaussian distributions. Left and right panels: influence of $\Gamma_0^{(1)}$ correction on one-shot $GW$ results with different starting mean-field calculations. Middle panel: improvement introduced by self-consistency upon one-shot $GW$ results.
}
\label{fig:error GW100} 
\end{figure*}

In Fig.~\ref{fig:Dev GW100}, we list the first IPs for the $GW$100 molecular data set.
In the top panel, {we compare the sc$GW$ and $G_0W_0\Gamma_0^{(1)}$ errors in IPs based on both the PBE and HF starting points.}
In the bottom panel, we compare IPs for sc$GW$ and  $G_0W_0$ method based on two starting points used in the top panel.
All the results are plotted as errors with respect to $\Delta$CCSD(T) reference values.~\cite{brunevalGWMiracleManyBody2021}

By categorizing the $GW$100 molecules into ten groups, we observe that sc$GW$ produces low and consistent {errors} for hydrides, halogenides, and most oxides. 
For dimers, hydrocarbons, and aromatics, the {errors} is higher. 
For compounds involving bonds with strong polar or ionic character (e.g., CF$_4$, SO$_2$, and MgO) sc$GW$ displays larger errors. 
Unsaturated bonding character (P$_2$, As$_2$, BN, HCN) also contributes to abnormal errors. 

In the bottom panel of Fig. \ref{fig:Dev GW100}, we observe that sc$GW$ improves one-shot $G_0W_0$ calculations. 
While the {MAE} of $G_0W_0$@HF is similar to the one of sc$GW$, we see that the majority of errors for sc$GW$ comes from nucleobases and MgO. In contrast,  $G_0W_0$@PBE displays many outliers for multiple system groups.

In the top panel of Fig.~\ref{fig:Dev GW100} and Fig.~\ref{fig:error GW100}, we observe that adding vertex corrections to $G_0W_0$@HF (denoted as $G_0 W_0 \Gamma_0^{(1)}$@HF) makes the results consistently worse, while adding vertex correction to $G_0W_0$@PBE  (denoted as $G_0 W_0 \Gamma_0^{(1)}$@PBE) improves the results, making the {MAE} comparable to sc$GW$.

{The effects of different starting points, PBE and HF, on $ \Gamma_0^{(1)}$ are analyzed in Fig.~\ref{fig:error GW100},} where we look at the trends in the error distribution curves for $G_0W_0$, $G_0W_0\Gamma_0^{(1)}$, and sc$GW$.
Regardless of the initial starting point used for $G_0W_0$, the vertex correction $\Gamma_0^{(1)}$ systematically enlarges the value of the first IP peak by a similar amount.
Because $G_0W_0$@PBE generally gives smaller IPs than $\Delta$CCSD(T) results, the uniform shift introduced by adding the vertex improves the 
 accuracy of $G_0 W_0 \Gamma_0^{(1)}$@PBE. 
For $G_0W_0$@HF, the IPs are already more accurate than those in $G_0W_0$@PBE. Consequently, adding vertex correction leads to diminished accuracy.
Based on this observation, we argue that the improvement of $\Gamma_0^{(1)}$ upon $G_0W_0$ is serendipitous and the accuracy of the overall result that includes vertex correction is mostly dictated by the starting point dependence.
Similar mean-field reference dependence of the vertex corrected results  has been observed for IPs of molecules containing transition metals.~\cite{wangVertexEffectsDescribing2022}
In sc$GW$, such a starting point dependence is effectively removed via the convergence of self-consistent loops.

In Table.~\ref{tab:mad_scgw}, we present statistical {criteria} for the data presented in Fig.~\ref{fig:Dev GW100}. We find that $G_0W_0\Gamma_0^{(1)}$@PBE gives lower {MAE} than sc$GW$ when compared with $\Delta$CCSD(T) benchmarks.
Even though sc$GW$ does not produce the best {MAE} out of all cases analyzed, one could still argue that it is more reliable to make predictions with the self-consistent scheme, because even with the pre-existing knowledge about the performance of $G_0W_0$ on a given system, it will still be difficult to know if $\Gamma_0^{(1)}$ would make improvement for such a system or not.

\begin{table}
{\centering
\begin{tabular}{ccccc}
\hline
\hline
 & $G_0W_0$ & $G_0W_0\Gamma_0^{(1)*}$ & sc$GW$ \\
\hline

@PBE & {0.62($\pm 0.29$)} & {0.20($\pm 0.26$)} & \multirow{2}{*}{{0.29($\pm 0.22$)}} \\ 
@HF  & {0.35($\pm 0.23$)} & {0.54($\pm 0.29$)} &  \\

\hline
\hline
\end{tabular}\par}
\caption{MAE in eV for $G_0W_0$, $G_0W_0\Gamma_0^{(1)}$, and sc$GW$ with $\Delta$CCSD(T) as benchmarks for the $GW$100 data set. 
The fluctuation in the bracket after each {MAE} value is the standard deviation of absolute errors. See Table SII in Supporting Information for detailed IP values for each molecule. $^*${Calculated with data} reported in Ref.~\cite{wangAssessingApproachHedin2021}.
}
\label{tab:mad_scgw}
\end{table}

\begin{table*}

{\centering
\begin{tabular}{ccccc}
\hline
\hline
 & $G_0W_0$@HF & $G_0W_0\Gamma^{\mathrm{(NL)}}$@HF & sc$GW$ & $\Delta$CCSD(T)$^*$ \\
\hline
cc-pVQZ & 0.65 ($\pm${0.36}) &  & 0.30 ($\pm${0.27}) & 0.23 ($\pm${0.32}) \\
cc basis limit & 0.88 ($\pm${0.38}) &  & 0.29 ($\pm${0.26}) &  \\
\hline
finite PW$^*$ & 0.42 ($\pm${0.37}) & 0.37 ($\pm${0.32}) &  &  \\
PW basis limit$^*$ & 0.69 ($\pm${0.40}) & 0.46 ($\pm${0.41}) &  & \\
\hline
\hline
\end{tabular}\par}
\caption{MAE in eV for $G_0W_0$@HF, sc$GW$, $G_0W_0\Gamma^{(\mathrm{NL})}$, and $\Delta$CCSD(T) methods for the 29-molecule data set, compared with experimental benchmarks. The fluctuation in the bracket after each MAE value is the standard deviation of absolute errors. See Table SIII in Supporting Information for detailed IP values for each molecule and the cited literature for experimental data. $^*$Calculated with data reported in Ref.~\cite{maggioGWVertexCorrected2017}.}
\label{tab:MAD_xperimental}
\end{table*}

\subsection{Comparison of sc$GW$ and $G_0W_0\Gamma$ with non-local vertex corrections for first IP peaks} \label{sec:Maggiovertex}

Maggio \textit{et al.} calculated first IP peaks for 29 molecules using their non-local vertex correction {$\Gamma^{(\mathrm{NL})}$}  on top of $G_0W_0$ in the plane wave (PW) basis.~\cite{maggioGWVertexCorrected2017}
Their first IP results were extrapolated to the PW basis set limit. 
Here, we compare our sc$GW$ and their {$G_0W_0\Gamma^{(\mathrm{NL})}$} for the first IP prediction against the experimental benchmarks (see Supporting Information Table SIII). 
{Note that while theoretical calculations assume vertical ionizations, this is not necessarily the case in experiments, where vibronic effects might apply}
Our $G_0W_0$ and sc$GW$ calculations are performed starting from HF/cc-pVXZ (X = Q and 5) and then extrapolated to the complete basis-set limit.
While we cannot compare these results in a very direct manner since they are performed in  different bases, we note that the PW basis and cc basis results behave similarly both in accuracy and basis set convergence trend, when compared against experimental benchmarks.
This is also confirmed by Maggio \textit{et al.} that GTO basis (cc-pVQZ) produced minimal numerical difference (about 100 meV) from finite PW results computed with the same $GW$ implementation.~\cite{maggioGWVertexCorrected2017}

In {Table} ~\ref{tab:MAD_xperimental}, for the first IPs, we summarize the overall {MAE} and standard deviation obtained using $G_0W_0$ and sc$GW$ and compare it against {$G_0W_0\Gamma^{(\mathrm{NL})}$} and $\Delta$CCSD(T) results from Ref.~\cite{maggioGWVertexCorrected2017}. 
Corresponding numerical data is presented in Table SIII of SI.
We observe that by extrapolating $G_0W_0$ and {$G_0W_0\Gamma^{(\mathrm{NL})}$} results from  a finite bases to their respective limits, the accuracy deteriorates even though higher number of orbitals are included. 
For cc basis sets, the {MAE} for our $G_0W_0$ increases from 0.65 eV (cc-pVQZ) to 0.88 eV (cc basis limit). 
Similarly, for the results in plane wave basis,~\cite{maggioGWVertexCorrected2017} the {MAE} of $G_0W_0$ increases from 0.42 eV (finite plane wave) to 0.69 eV (plane wave basis limit),
and the {MAE} of {$G_0W_0\Gamma^{(\mathrm{NL})}$} increases from 0.37 eV (finite plane wave) to {0.46} eV (plane wave basis limit). 
In Ref. \cite{maggioGWVertexCorrected2017}, the extrapolated {$G_0W_0\Gamma^{(\mathrm{NL})}$} results were also coupled with convergence issue for some molecules (lithium dimer, phosphorus dimer, and sulfur dioxide).
On the other hand sc$GW$ results are essentially converged already at the cc-pVQZ level. {As a result, for sc$GW$, cc-pVQZ results and cc basis limit results are close for most entries (see Supporting Information Table SIII).}

\begin{figure*}
\includegraphics[scale=0.70]{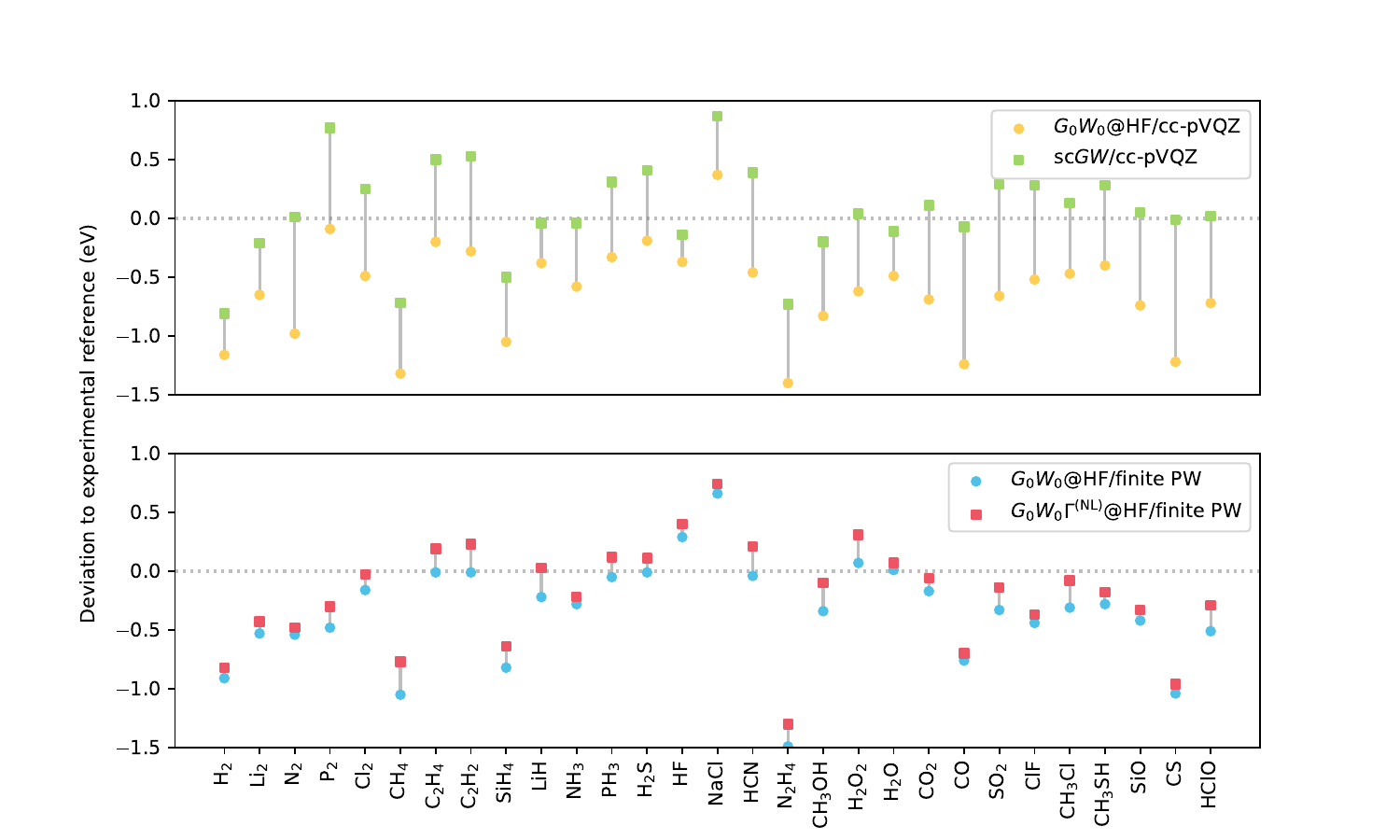}
\centering
\caption{Signed errors from the experimental data, of $G_0W_0${@HF}, sc$GW$, and {$G_0W_0\Gamma^{(\mathrm{NL})}$} for the 29-molecule data set. Top panel: improvement brought upon $G_0 W_0$ prediction by self-consistency in a finite cc basis. Bottom panel: improvement brought by the vertex correction on top of $G_0 W_0$ in a finite plane wave basis, calculated with data reported in Ref.~\cite{maggioGWVertexCorrected2017}.}
\label{fig:K29 dev} 
\end{figure*}

Table~\ref{tab:MAD_xperimental} shows that in the CBS limit for plane waves, vertex correction reduces the {MAE} for the IPs from 0.69 eV in $G_0W_0$ to 0.46 eV.
Meanwhile, in the CBS limit for the cc basis set family, self-consistency improves the {MAE} from 0.88 eV at the $G_0W_0$ level to 0.29 eV.

The magnitude of improvement induced by both self-consistency and vertex correction is illustrated in Fig.~\ref{fig:K29 dev}. In the top panel of Fig.~\ref{fig:K29 dev}, we observe that self-consistency in most cases brings the IP values closer to experiment in comparison to $G_0 W_0$. 
The bottom panel shows the magnitude of  the improvement brought by vertex correction on top of $G_0 W_0$, which is only minor. Thus, its final accuracy largely depends on a good mean-field starting pointing point of the $G_0 W_0$ calculations. 

Overall, we conclude that in the molecular IP domain, improvements introduced by self-consistency is similar to, if not better than, that of vertex corrected $G_0W_0$. 
{Moreover, the smaller standard deviation in sc$GW$ implies more uniformity in the quality of results.}
This, combined with the lack of dependence on the starting mean-field solution, makes sc$GW$ more favorable.

\section{Conclusion}\label{sec:conclusion}

In this work, we demonstrated the performance of our finite temperature sc$GW$ methodology in predicting molecular valence shell IPs. In our implementation, there are no approximations other than (a) density fitting approximation for integral generation; and (b) the Nevanlinna analytical continuation employed to obtain spectral data from the converged Green's function evaluated on the imaginary axis.

Based on our calculations, performing the self-consistency generally improves upon $G_0W_0$ results and leads to convergence of calculations with different mean-field starting points.
This eliminates the ambiguity associated with the selecting a mean-field calculation used as the reference for the $GW$ method.
The reliability of our sc$GW$ method is verified both for the first IPs as well as the inner valence shell IPs, when examined against both theoretical and experimental benchmarks. 

For molecular systems, we  presented comparisons of our sc$GW$ with another post-$GW$ methodology -- vertex corrected $G_0W_0$ -- motivated by completing the suite of Hedin's equations. 
When comparing different $G_0W_0\Gamma$ variants against sc$GW$, we observed that sc$GW$ consistently displays either better or comparable accuracy. 
The $G_0W_0\Gamma$ results were affected by a strong starting point dependence (inherited from $G_0W_0$) and the magnitude of error caused by starting points is frequently larger than the correction introduced by the vertex. Similar dependence was also observed in vertex correction upon polarizability exclusively.~\cite{lewisVertexCorrectionsPolarizability2019}

Even though there are scattered cases where $G_0W_0\Gamma$ based on a DFT reference outperforms sc$GW$, full self-consistency is cheaper than evaluating full vertex corrections and gives unbiased results independent of the starting point. Moreover, within the sc$GW$ framework, the evaluation of total energies and, consequently, energy differences 
is possible.~\cite{galitskii1958application,holmTotalEnergyGalitskiiMigdal2000,stanLevelsSelfconsistencyGW2009,carusoSelfconsistentGWAllelectron2013a} {In contrast, in the $G_0 W_0$ schemes, the energy is ill defined since its value strongly depends on the underlying reference. Additionally, the self-consistency in sc$GW$ causes relaxation of the orbitals in the presence of correlation resulting in improved quantities such as the electron density.~\cite{carusoUnifiedDescriptionGround2012,carusoSelfconsistentGWAllelectron2013a, carusoSelfconsistentGWAllelectron2013b} }

Moreover, choosing an appropriate type of vertex correction can be a difficult task. In this work, we analyzed three different versions of vertex corrections each employing different approximations and we concluded that it is hard to establish \textit{a priori} which type of vertex correction should be used for a certain problem. 

{While sc$GW$ is generally accurate for the first IP as well as the inner valence shell excitations, depiction of excitations with MQP character is believed to be relatively difficult
due to their correlation effects. 
Nevertheless, at least for the examples analyzed in the Results section, it appears that sc$GW$ is capable of not only capturing the qualitative emergence of MQP features but also yielding reasonably accurate excitation energies in this regime.}
This is particularly advantageous in comparison to methods such as EOM-CCSD, where the presence of satellites may lead to issues with converging the quasiparticle energies.

In summary, vertex correction is generally considered as a preferred way to improve the quality of $G_0W_0$ results.
However, we find that at least for molecules, sc$GW$, without vertex, already provides results that are competitive with the best $G_0W_0 \Gamma$ results analyzed here.
This, combined with the fact that sc$GW$ is essentially a black-box method, makes self-consistency a better route to make improvements upon $G_0W_0$.

In the future, one still may want to add additional diagrammatic terms beyond sc$GW$. 
The best way of doing it, however, is an active field of research. 
Ideally, the implementation of vertex corrections should be done based on the self-consistent $GW$ approach where the starting point dependence is removed. 
Only then the approximations introduced in the formulation of $GW\Gamma$ can be meaningfully validated. 
The self-consistent evaluation of Hedin's equations with an addition of the vertex may {become} numerically difficult and may result in the appearance of unphysical features such as negative spectral functions.~\cite{pavlyukhDynamicallyScreenedVertex2020a} 
{When applied within the self-consistent $GW$ scheme, the vertex correction} is responsible only for bringing the correlation that is missing in the parent sc$GW$ approach, and it does not need to remedy for the lack of the orbital optimization that is present in $G_0W_0$.
This direction will be explored in our future work.

\begin{acknowledgement}
M.W., G.H., V.A., A.S., {K.B.W.} and D.Z. are supported by the U.S. Department of Energy, Office of Science, Office of Advanced Scientific Computing
Research and Office of Basic Energy Sciences, Scientific
Discovery through Advanced Computing (SciDAC) program under Award Number DE-SC0022198.
\end{acknowledgement}

\begin{suppinfo}

Detailed ionization potential values, spectral functions, and additional comparisons between sc$GW$ and $G_0W_0$. 

\end{suppinfo}

\bibliography{Molecules_Gamma_Uniq}



\section*{Supplementary material}

# Supporting Information:

# Comparing self-consistent $GW$ and vertex corrected $G_0W_0$ ($G_0W_0\Gamma$) accuracy for molecular ionization potentials

Ming Wen,[†] Vibin Abraham,[†] Gaurav Harsha,[†] Avijit Shee,[‡] K. Birgitta Whaley,[‡] and Dominika Zgid[*,†,¶]

[†]Department of Chemistry, University of Michigan, Ann Arbor, Michigan 48109

[‡]Department of Chemistry, University of California, Berkeley, California 94720-1460

[¶]Department of Physics, University of Michigan, Ann Arbor, Michigan 48109

E-mail: zgid.umich@edu

# A. Inner valence shell ionizations

In Table S1, we report our detailed data for the first and few inner valence ionization peaks for each molecule entry. For ammonia, water, nitrogen, acetylene, and methane, $G_0W_0\Gamma_X$ and ASCI results were reported by Mejuto-Zaera *et. al.*[1] For $G_0W_0$, $G_0W_0$, sc$GW$, and ASCI, results are extrapolated to the aug-cc basis limit. The EOM columns are computed using aug-cc-pVQZ and not extrapolated.

Table S1: First few ionization potential predictions (in eV) for molecules. $G_0W_0$, $G_0W_0\Gamma$, and ASCI data cited from Ref.[1] are reproduced in this table with permission, © Copyright 2024 AIP Publishing LLC. *Methane's $G_0W_0$, $G_0W_0\Gamma$, and ASCI results were reported in cc basis. **Unable to converge.

| Molecule | Peak | $G_0W_0$[1] | $G_0W_0\Gamma_X$[1] | sc$GW$ | EOM-CCSD | EOM-CCSDT | ASCI[1] |
|---|---|---|---|---|---|---|---|
| Ethylene | 1 | | | -10.28 | -10.76 | -10.72 | |
| | 2 | | | -13.20 | -13.16 | -13.10 | |
| | 3 | | | -14.83 | -15.01 | -14.86 | |
| | 4 | | | -16.36 | -16.38 | -16.20 | |
| Hydrogen fluoride | 1-2 | | | -16.39 | -16.15 | -16.10 | |
| | 3 | | | -20.09 | -20.06 | -20.01 | |
| | 4 | | | -39.57 | ** | ** | |
| Carbon monoxide | 1 | | | -14.20 | -14.50 | -14.15 | |
| | 2-3 | | | -15.20 | -15.63 | -15.57 | |
| | 4 | | | -19.64 | -19.57 | -19.02 | |
| | 5 | | | -33.57 | ** | ** | |
| Ammonia | 1 | -11.70 | -10.85 | -11.00 | -10.94 | -10.93 | -10.81 |
| | 2-3 | -17.12 | -16.70 | -16.58 | -16.65 | -16.61 | -16.51 |
| | 4 | -31.08 | -27.58 | -28.07 | -27.90 | -26.97 | -27.36 |
| Water | 1 | -13.36 | -12.94 | -12.84 | -12.70 | -12.70 | -12.75 |
| | 2 | -15.44 | -15.03 | -14.99 | -14.90 | -14.91 | -14.90 |
| | 3 | -19.62 | -19.20 | -19.02 | -19.07 | -19.06 | -19.10 |
| | 4 | -35.06 | -32.02 | -32.91 | -32.88 | -32.73 | -32.99 |
| Nitrogen | 1 | -17.28 | -16.28 | -15.69 | -15.74 | -15.60 | -15.54 |
| | 2-3 | -16.82 | -17.11 | -16.48 | -17.34 | -17.05 | -17.05 |
| | 4 | -21.14 | -19.62 | -19.14 | ** | -18.98 | -18.88 |
| Acetylene | 1-2 | -11.18 | -11.27 | -11.05 | -11.66 | -11.51 | -11.45 |
| | 3 | -18.53 | -17.95 | -17.25 | -17.34 | -17.19 | -17.15 |
| | 4 | -20.90 | -19.62 | -19.32 | -19.21 | -19.05 | -19.05 |
| | 5 | -28.06 | -24.31 | -24.65 | -24.57 | -24.70 | -24.13 |
| Methane* | 1-3 | -14.83 | -14.61 | -14.39 | -14.45 | -14.36 | -14.35 |
| | 4 | -25.69 | -22.74 | -23.43 | -23.42 | -23.12 | -23.25 |

In Fig. S1 and S2, we present five examples of spectral functions produced from the

sc$GW$/aug-cc-pVQZ calculations via Nevanlinna analytical continuation. Similar spectral functions rendered by stochastic $G_0W_0\Gamma_X$ were reported in Ref.[1] Notice that in ammonia and water we had the complicated multi-peak feature in the MQP regime. We identify the MQP peak by increasing the broadening factor so that the complicated multi-peak feature melted into a singular broad peak, which corresponds to the IP number we report in TABLE S1.

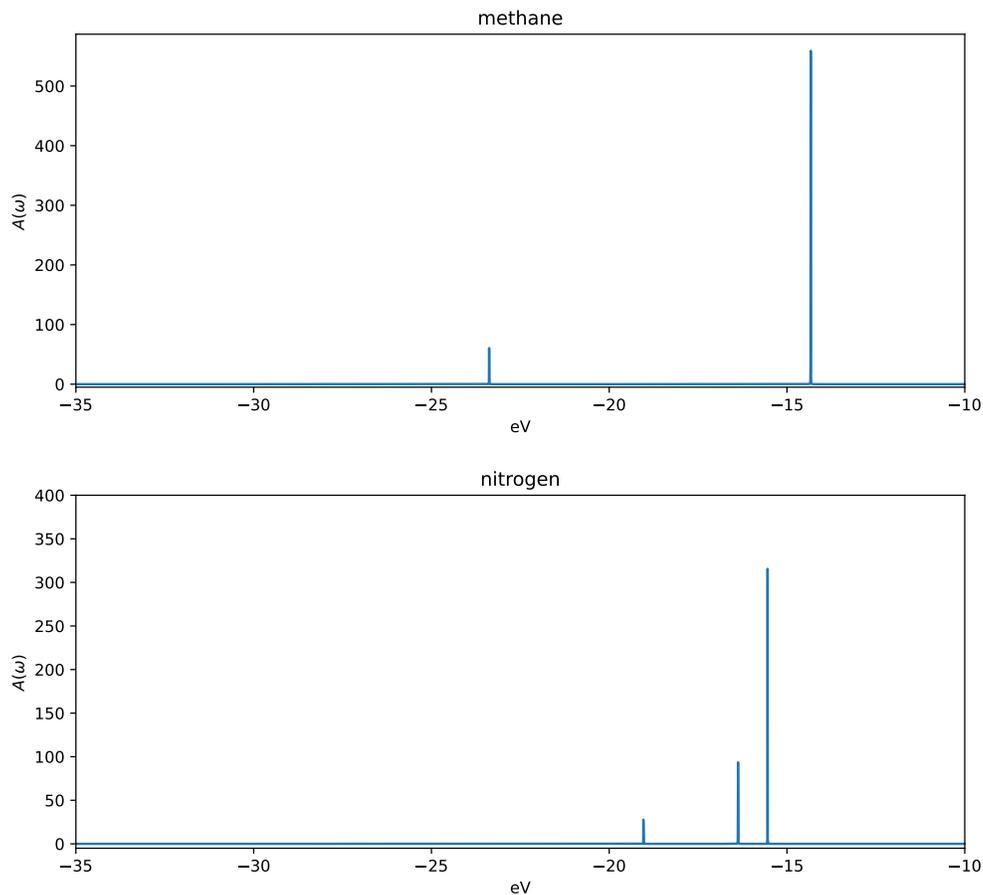

Figure S1: Spectral functions of methane and nitrogen. The spectral functions were produced using the analytically continued data from sc$GW$ results.

In Fig. S3, we present the absolute deviation when changing the ASCI benchmarks[1] to EOM-CCSDT benchmarks.

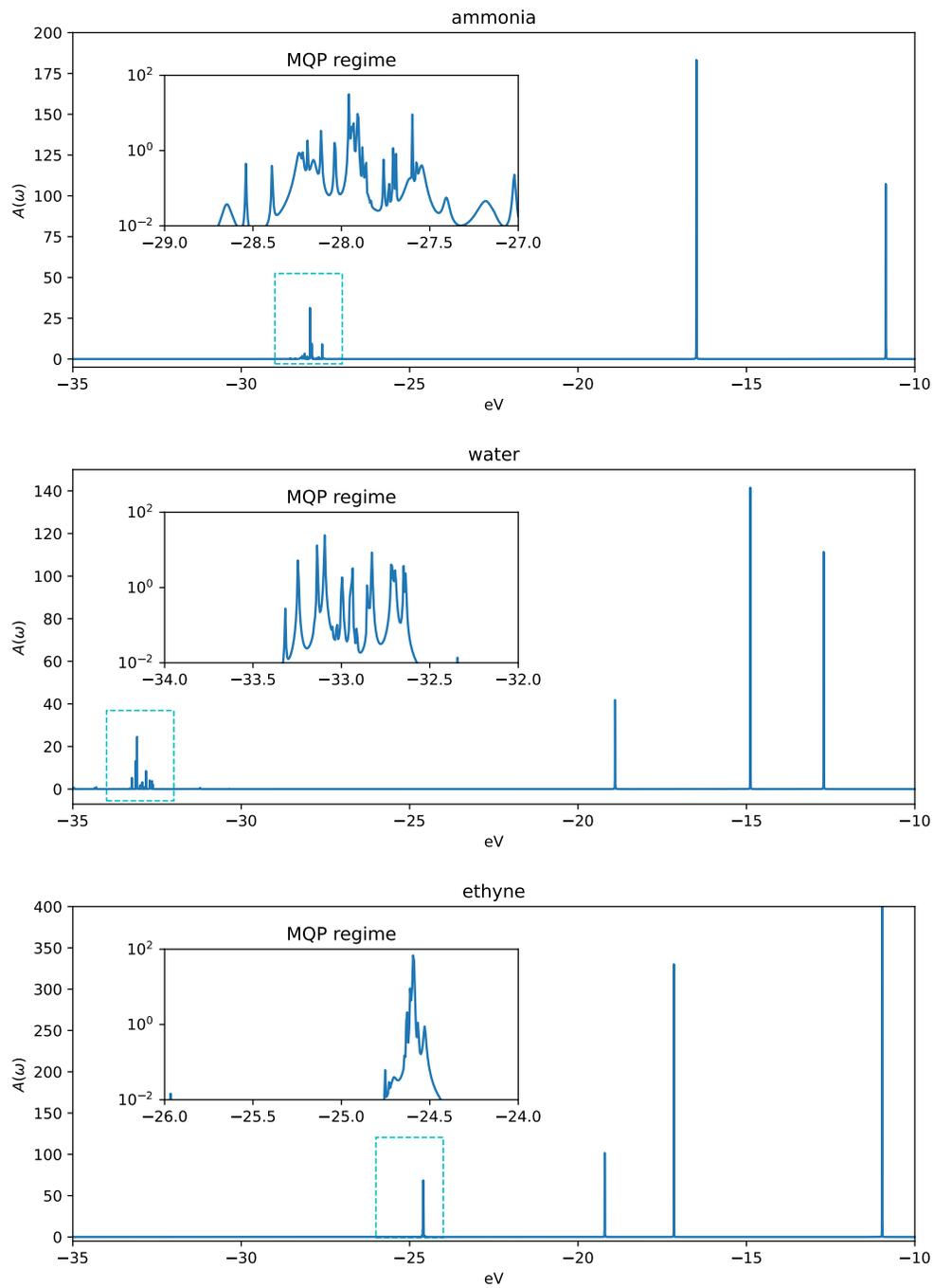

Figure S2: Spectral functions of ammonia, water, and acetylene. The spectral functions were produced using the analytically continued data from sc$GW$ results. The MQP regimes insets are plotted in algorithmic scale.

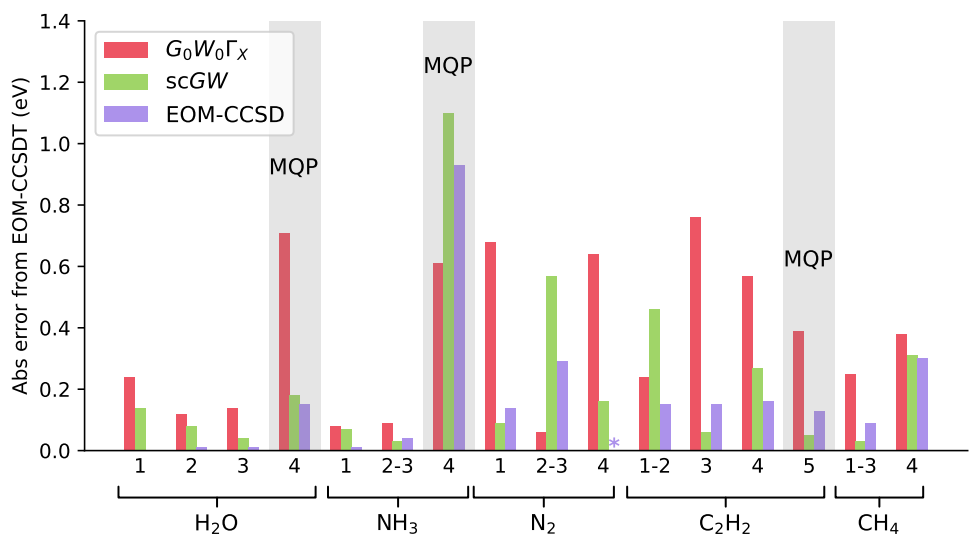

Figure S3: Absolute error of sc$GW$, $G_0W_0\Gamma_X$, and EOM-CCSD predictions of the first and inner shell ionization peaks from the EOM-CCSDT data. *EOM-CCSD calculation did not converge.

## B. First ionization peaks

We present our validation test on the $GW100$ set. We compared our sc$GW$ method to $G_0W_0\Gamma_0^{(1)}$ by Wang *et al.*[2] while keeping the level of theory as similar as possible. Initializations for both approaches are done using HF/def2-TZVPP and PBE/def2-TZVPP. Molecules containing elements in the fourth row or beyond are excluded, since the all-electron basis set of def2-TZVPP for these elements was not available. Total of 93 molecules are kept.

Firstly, we need to confirm that finite temperature sc$GW$ (on the imaginary axis) and $G_0W_0\Gamma_0^{(1)}$ (on the real axis) can produce the same results after the first iteration, *i.e.* the $G_0W_0$ level. If so, we can make sure the comparison is appropriate and based on similar theoretical footing. In Fig. S4 the $G_0W_0$ results for both schemes fit well with no obvious outliers. Additionally, we compared our sc$GW$ with past implementation of sc$GW$ by Caruso *et al.*[3] This observation also supports that one can reliably recover spectral information from the Green's function on imaginary axis via analytical continuation.

Table S2: First ionization potentials (in eV) calculated from $G_0W_0$@HF, $G_0W_0$@PBE, sc$GW$, and $\Delta$CCSD(T) for the $GW$100 set.

| Index | Molecule | $G_0W_0$@HF | $G_0W_0$@PBE | sc$GW$ | CCSD(T)[4] | Index | Molecule | $G_0W_0$@HF | $G_0W_0$@PBE | sc$GW$ | CCSD(T)[4] |
|---|---|---|---|---|---|---|---|---|---|---|---|
| 1 | He | -24.58 | -23.75 | -24.41 | -24.51 | 53 | HCl | -12.81 | -12.13 | -12.28 | -12.59 |
| 2 | Ne | -21.39 | -20.49 | -21.39 | -21.32 | 54 | LiF | -11.37 | -9.77 | -11.36 | -11.32 |
| 3 | Ar | -15.76 | -15.02 | -15.26 | -15.54 | 55 | $MgF_2$ | -13.79 | -12.29 | -13.77 | -13.71 |
| 4 | Kr | -14.00 | -13.37 | -13.64 | -13.94 | 56 | $TiF_4$ | -16.17 | -13.77 | -15.47 | -15.41 |
| 6 | $H_2$ | -16.48 | -15.90 | -16.15 | -16.40 | 57 | $AlF_3$ | -15.63 | -14.14 | -15.41 | -15.32 |
| 7 | $Li_2$ | -5.32 | -5.05 | -4.97 | -5.27 | 58 | BF | -11.31 | -10.43 | -10.61 | -11.09 |
| 8 | $Na_2$ | -4.97 | -4.92 | -4.63 | -4.95 | 59 | $SF_4$ | -13.30 | -11.91 | -12.49 | -12.59 |
| 9 | $Na_4$ | -4.29 | -4.11 | -3.83 | -4.24 | 60 | KBr | -8.21 | -7.15 | -7.88 | -8.13 |
| 10 | $Na_6$ | -4.47 | -4.25 | -3.96 | -4.39 | 61 | GaCl | -9.90 | -9.42 | -9.34 | -9.77 |
| 11 | $K_2$ | -4.06 | -3.94 | -3.73 | -4.07 | 62 | NaCl | -9.24 | -7.90 | -8.77 | -9.03 |
| 13 | $N_2$ | -16.35 | -14.74 | -15.45 | -15.48 | 63 | $MgCl_2$ | -11.93 | -10.84 | -11.44 | -11.66 |
| 14 | $P_2$ | -10.55 | -10.05 | -9.73 | -10.53 | 65 | BN | -11.76 | -11.08 | -11.03 | -11.98 |
| 15 | $As_2$ | -9.74 | -9.34 | -9.05 | -9.85 | 66 | HCN | -13.87 | -13.00 | -13.13 | -13.72 |
| 16 | $F_2$ | -16.31 | -14.83 | -15.80 | -15.58 | 67 | PN | -12.37 | -10.99 | -11.59 | -11.81 |
| 17 | $Cl_2$ | -11.77 | -10.90 | -11.08 | -11.41 | 68 | $N_2H_4$ | -10.17 | -9.21 | -9.63 | -9.69 |
| 18 | $Br_2$ | -10.75 | -9.98 | -10.21 | -10.52 | 69 | $CH_2O$ | -11.37 | -10.17 | -10.82 | -10.84 |
| 20 | $CH_4$ | -14.77 | -13.91 | -14.25 | -14.37 | 70 | $CH_3OH$ | -11.57 | -10.46 | -11.04 | -11.04 |
| 21 | $C_2H_6$ | -13.21 | -12.35 | -12.63 | -12.71 | 71 | EtOH | -11.29 | -10.05 | -10.67 | -10.69 |
| 22 | $C_3H_8$ | -12.63 | -11.66 | -12.03 | -12.03 | 72 | $CH_3CHO$ | -10.81 | -9.40 | -10.18 | -10.21 |
| 23 | $C_4H_{10}$ | -12.19 | -11.23 | -11.65 | -11.57 | 73 | $Et_2O$ | -10.49 | -9.21 | -9.81 | -9.82 |
| 24 | $C_2H_4$ | -10.71 | -10.20 | -10.11 | -10.67 | 74 | HCOOH | -11.95 | -10.59 | -11.41 | -11.42 |
| 25 | $C_2H_2$ | -11.59 | -10.94 | -10.86 | -11.42 | 75 | $H_2O_2$ | -12.06 | -10.87 | -11.54 | -11.52 |
| 26 | $C_4$ | -11.62 | -10.64 | -10.69 | -11.24 | 76 | $H_2O$ | -12.87 | -11.94 | -12.57 | -12.57 |
| 27 | $C_3H_6$ | -11.30 | -10.46 | -10.59 | -10.87 | 77 | $CO_2$ | -14.21 | -13.07 | -13.54 | -13.71 |
| 28 | $C_6H_6$ | -9.52 | -8.87 | -8.75 | -9.36 | 78 | $CS_2$ | -10.33 | -9.55 | -9.44 | -9.98 |
| 29 | $C_8H_8$ | -8.67 | -7.93 | -7.82 | -8.40 | 79 | OCS | -11.55 | -10.74 | -10.70 | -11.17 |
| 30 | $C_5H_6$ | -8.86 | -8.23 | -8.07 | -8.71 | 80 | COSe | -10.69 | -10.00 | -9.98 | -10.47 |
| 31 | $C_2H_3F$ | -10.79 | -10.02 | -10.08 | -10.55 | 81 | CO | -15.06 | -13.43 | -13.97 | -14.21 |
| 32 | $C_2H_3Cl$ | -10.34 | -9.58 | -9.61 | -10.09 | 82 | $O_3$ | -13.54 | -11.73 | -12.57 | -12.71 |
| 33 | $C_2H_3Br$ | -9.46 | -8.83 | -8.84 | -9.27 | 83 | $SO_2$ | -12.94 | -11.61 | -12.03 | -12.30 |
| 35 | $CF_4$ | -16.85 | -15.18 | -16.55 | -16.23 | 84 | BeO | -9.83 | -9.16 | -9.75 | -9.94 |
| 36 | $CCl_4$ | -12.01 | -10.77 | -11.14 | -11.50 | 85 | MgO | -7.95 | -7.05 | -7.95 | -7.91 |
| 37 | $CBr_4$ | -10.78 | -9.67 | -10.12 | -10.41 | 86 | $C_6H_5CH_3$ | -9.16 | -8.49 | -8.37 | -8.97 |
| 39 | $SiH_4$ | -13.25 | -12.28 | -12.73 | -12.80 | 87 | $C_6H_5Et$ | -9.13 | -8.43 | -8.32 | -8.92 |
| 40 | $GeH_4$ | -12.88 | -11.99 | -12.37 | -12.50 | 88 | $C_6F_6$ | -10.63 | -9.28 | -9.50 | -9.93 |
| 41 | $Si_2H_6$ | -11.11 | -10.21 | -10.44 | -10.65 | 89 | $C_6H_5OH$ | -9.03 | -8.22 | -8.17 | -8.70 |
| 42 | $Si_5H_{12}$ | -9.82 | -8.81 | -9.04 | -9.27 | 90 | $C_6H_5NH_2$ | -8.35 | -7.49 | -7.52 | -8.04 |
| 43 | LiH | -8.17 | -7.02 | -7.89 | -7.96 | 91 | $C_5H_5N$ | -9.91 | -8.87 | -9.12 | -9.73 |
| 44 | KH | -6.29 | -4.81 | -6.03 | -6.13 | 92 | guanine | -8.44 | -7.52 | -7.48 | -8.03 |
| 45 | $BH_3$ | -13.67 | -12.84 | -13.18 | -13.27 | 93 | adenine | -8.71 | -7.80 | -7.78 | -8.33 |
| 46 | $B_2H_6$ | -12.76 | -11.74 | -12.17 | -12.25 | 94 | cytosine | -9.28 | -8.08 | -8.40 | -8.77 |
| 47 | $NH_3$ | -11.19 | -10.29 | -10.77 | -10.81 | 95 | thymine | -9.68 | -8.49 | -8.70 | -9.08 |
| 48 | $HN_3$ | -11.11 | -10.27 | -10.25 | -10.68 | 96 | uracil | -10.09 | -8.86 | -9.13 | -9.48 |
| 49 | $PH_3$ | -10.81 | -10.20 | -10.23 | -10.52 | 97 | urea | -10.68 | -9.18 | -10.05 | -10.05 |
| 50 | $AsH_3$ | -10.58 | -10.04 | -10.08 | -10.40 | 99 | $Cu_2$ | -7.20 | -7.61 | -6.96 | -7.57 |
| 51 | $H_2S$ | -10.52 | -9.92 | -9.99 | -10.31 | 100 | CuCN | -11.29 | -9.80 | -10.67 | -10.85 |
| 52 | HF | -16.22 | -15.29 | -16.13 | -16.03 | | MAE | 0.32 | 0.62 | 0.29 | |

7

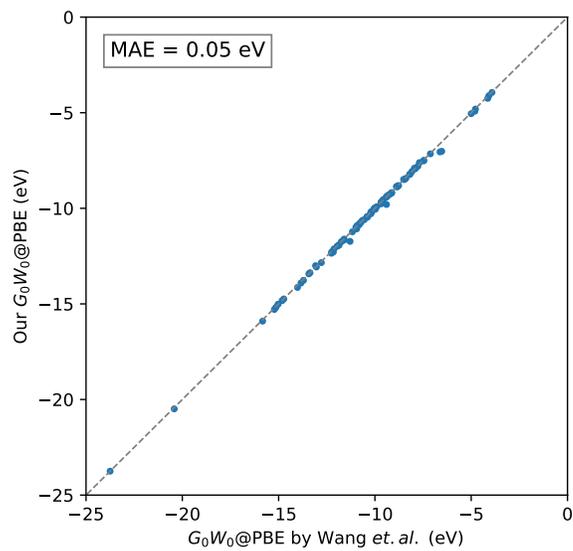

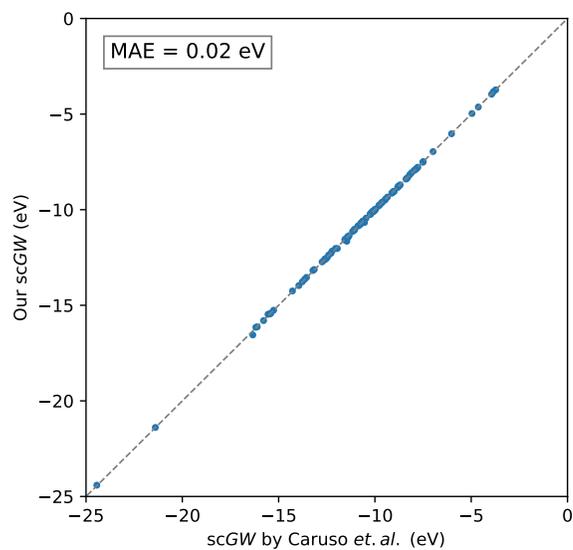

Figure S4: Left: Comparison of ionization potentials derived from $G_0W_0$@PBE by Wang *et al.* and our finite temperature version of $G_0W_0$@PBE. Right: Comparison of ionization potentials derived from sc$GW$ by Caruso *et al.* and our finite temperature version of sc$GW$.

8

We use HF/cc-pVXZ (X = Q and 5) as the starting point for our subsequent scGW calculations for the 29-molecule data set. Then we compare these results with the vertexed corrected $G_0W_0$ method ($G_0W_0\Gamma^{(NL)}$ in the main text) proposed by Maggio *et. al.*[5] Final sc$GW$ results are linearly extrapolated to cc basis set limit. The cited data calculated in the plane wave basis were also evaluated with HF starting mean field. Please refer to the original article for their detailed description for energy cut-offs and the extrapolation technique for the PW basis.

Table S3: First ionization potentials (in eV) calculated from CCSD(T), $G_0W_0$@HF, sc$GW$, and $G_0W_0\Gamma$@HF, in cc basis and plane wave basis sets. *Vertical IP values are in italics.

| | cc-pVQZ | | | cc basis limit | | finite PW[5] | | PW basis limit[5] | | |
|---|---|---|---|---|---|---|---|---|---|---|
| Molecule | CCSD(T)[5] | $G_0W_0$ | sc$GW$ | $G_0W_0$ | sc$GW$ | $G_0W_0$ | $G_0W_0\Gamma$ | $G_0W_0$ | $G_0W_0\Gamma$ | Experiment* |
| hydrogen | -16.39 | -16.59 | -16.24 | -16.65 | -16.32 | -16.34 | -16.25 | -16.72 | -16.52 | -15.43[6] |
| lithium dimer | -5.17 | -5.38 | -4.94 | -5.42 | -4.94 | -5.26 | -5.16 | -5.29 | | -4.73[7] |
| nitrogen | -15.49 | -16.56 | -15.57 | -16.82 | -15.66 | -16.12 | -16.06 | -16.56 | -16.39 | -15.58[8] |
| phosphorus dimer | -10.76 | -10.71 | -9.85 | -11.01 | -9.98 | -11.10 | -10.92 | -11.31 | | -10.62[9] |
| chlorine | -11.62 | -11.98 | -11.24 | -12.35 | -11.37 | -11.65 | -11.52 | -12.08 | -11.80 | *-11.49*[10] |
| methane | -14.40 | -14.92 | -14.32 | -15.05 | -14.37 | -14.65 | -14.37 | -14.95 | -14.57 | *-13.6*[11] |
| ethylene | -10.69 | -10.88 | -10.18 | -11.04 | -10.24 | -10.69 | -10.49 | -10.91 | -10.66 | *-10.68*[11] |
| acetylene | -11.42 | -11.77 | -10.96 | -11.95 | -11.13 | -11.50 | -11.26 | -11.73 | -11.43 | *-11.49*[11] |
| silane | -12.82 | -13.35 | -12.80 | -13.50 | -12.89 | -13.12 | -12.94 | -13.40 | -12.88 | *-12.3*[12] |
| lithium hydride | -7.94 | -8.28 | -7.94 | -8.39 | -8.02 | -8.12 | -7.87 | -8.26 | -7.94 | -7.9[13] |
| ammonia | -10.92 | -11.40 | -10.86 | -11.60 | -10.98 | -11.10 | -11.04 | -11.45 | -11.28 | -10.82[14] |
| phosphine | -10.49 | -10.92 | -10.28 | -11.14 | -10.48 | -10.64 | -10.47 | -10.89 | -10.66 | -10.59[15] |
| hydrogen sulfide | -10.43 | -10.69 | -10.09 | -10.97 | -10.15 | -10.51 | -10.39 | -10.79 | -10.60 | -10.50[16] |
| hydrogen fluoride | -16.09 | -16.49 | -16.26 | -16.79 | -16.41 | -15.83 | -15.72 | -16.29 | -16.18 | -16.12[17] |
| sodium chloride | -9.13 | -9.43 | -8.93 | -9.83 | -9.09 | -9.14 | -9.06 | -9.51 | -9.32 | *-9.80*[18] |
| hydrogen cyanide | -13.64 | -14.07 | -13.22 | -14.28 | -13.44 | -13.65 | -13.4 | -13.92 | -13.61 | *-13.61*[19] |
| hydrazine | -10.24 | -10.38 | -9.71 | -10.61 | -9.83 | -10.47 | -10.28 | -10.86 | -10.58 | *-8.98*[20] |
| methanol | -11.08 | -11.79 | -11.16 | -12.03 | -11.33 | -11.30 | -11.06 | -11.71 | -11.39 | *-10.96*[21] |
| hydrogen peroxide | -11.49 | -12.32 | -11.66 | -12.62 | -11.91 | -11.63 | -11.39 | -12.12 | -11.81 | *-11.70*[22] |
| water | -12.64 | -13.11 | -12.73 | -13.38 | -12.80 | -12.61 | -12.55 | -13.10 | -12.92 | *-12.62*[23] |
| carbon dioxide | -13.78 | -14.46 | -13.66 | -14.78 | -13.92 | -13.94 | -13.83 | -14.35 | -14.16 | *-13.77*[24] |
| carbon monoxide | -14.05 | -15.25 | -14.08 | -15.47 | -14.17 | -14.77 | -14.71 | -15.03 | -14.89 | *-14.01*[25] |
| sulfur dioxide | -12.41 | -13.16 | -12.21 | -13.60 | -12.48 | -12.83 | -12.64 | -13.20 | | *-12.50*[23] |
| chlorine fluoride | -12.82 | -13.29 | -12.49 | -13.67 | -12.73 | -13.21 | -13.14 | -13.49 | -13.33 | *-12.77*[10] |
| chloromethane | -11.41 | -11.76 | -11.16 | -12.10 | -11.24 | -11.60 | -11.37 | -11.90 | -11.56 | *-11.29*[23] |
| methanethiol | -9.49 | -9.84 | -9.16 | -10.12 | -9.35 | -9.72 | -9.62 | -9.93 | -9.76 | *-9.44*[26] |
| silicon monoxide | -11.55 | -12.04 | -11.25 | -12.30 | -11.35 | -11.72 | -11.63 | -12.05 | -11.78 | -11.3[27] |
| carbon monosulfide | -11.45 | -12.55 | -11.34 | -12.80 | -11.44 | -12.37 | -12.29 | -12.63 | -12.47 | -11.33[28] |
| hypochlorous acid | -11.30 | -11.84 | -11.10 | -12.20 | -11.30 | -11.63 | -11.41 | -11.96 | -11.66 | -11.12[29] |
| MAE from experiment | 0.23 | 0.65 | 0.30 | 0.88 | 0.29 | 0.42 | 0.37 | 0.69 | 0.46 | |

9

 

\end{document}